\documentclass[article]{aa}

\usepackage{float}
\usepackage{graphicx}
\usepackage{placeins}
\usepackage{aas_macros}
\usepackage{comment}

\usepackage{txfonts}

\usepackage{hyperref}
\hypersetup{
    colorlinks,
    citecolor=blue,
    filecolor=blue,
    linkcolor=blue,
    urlcolor=blue
}

\newcommand{\rev}[1]{#1}

\begin{document} 

\title{Stellar age determination using deep neural networks}
\subtitle{Isochrone ages for 1.3 million stars, based on BaSTI, MIST, PARSEC, Dartmouth, and SYCLIST evolutionary grids}
\titlerunning{Stellar age determination using deep neural networks}

\author{T. Boin\inst{1}\and
      L. Casamiquela\inst{1}\and
      M. Haywood\inst{1}\and
      P. Di Matteo\inst{1}\and
      Y. Lebreton\inst{1,2} \and
      M. Uddin\inst{1} \and
      D. R. Reese\inst{1}
      }

\institute{LIRA, Observatoire de Paris, PSL Research University, CNRS, Univ Paris Diderot, Sorbonne Paris Cité, Place Jules Janssen, 92195, Meudon, France\\\email{tristan.boin@obspm.fr}
\and 
Univ Rennes, CNRS, IPR (Institut de Physique de Rennes) – UMR 6251, 35000 Rennes, France}

\date{Received XXX; accepted XXX}

\abstract{
Recent spectroscopic surveys provide element abundances for large samples of Milky Way stars, from which stellar parameters can be inferred. Stellar ages, among them, are both a notoriously difficult parameter to estimate and a fundamental property for Galactic archaeology studies.
}{
We aim to develop a model-driven deep learning approach to age determination by training neural networks on stellar evolutionary grids. Contrary to the usual data-driven deep learning approach of using prior age estimates as training data, our method has the potential for a wider and less biased range of application. The low computational cost of deep learning methods compared to, for example, Bayesian isochrone fitting enables a broad analysis of large spectroscopic catalogues.
}{
We trained multilayer perceptrons on different stellar evolutionary grids to map [M/H], $M_G$, $(G_{BP}-G_{RP})$ to stellar age $\tau$. We combined $Gaia$ photometry and parallaxes, metallicities, and $\alpha$ elements from spectroscopic surveys and extinction maps, which are passed through neural networks to estimate stellar ages.
}{
We applied our method to the LAMOST DR10, GALAH DR3 \& DR4, and APOGEE DR17 spectroscopic surveys, estimating ages using the BaSTI tracks and other stellar evolutionary models. We leveraged this novel technique to study, for the first time, differences in age estimates from several evolutionary grids applied to very large datasets. In addition, we dated 13 open clusters and one globular cluster, finding a median absolute deviation with literature ages of 0.20 Gyr. Along with the stellar age catalogues from our estimates, we release \texttt{NEST} (Neural Estimator of Stellar Times), a python package to estimate stellar age based on this work, as well as a web interface.
}
{
We show that, when using the same evolutionary grid, our method retrieves the same ages as a Bayesian approach similar to SPInS, for only a fraction of the computational cost, with a 60,000 speed-up factor for a typical star. This model-driven deep learning technique thus opens up the way for broad galactic archaeology studies on the largest datasets available today and in the near future with upcoming surveys such as 4MOST.
}
\maketitle

\section{Introduction}

Galactic archaeology relies on a combination of precise abundance measurements and age estimates to study stellar populations and their evolution. On the chemical abundances side, large spectroscopic surveys such as the Large Sky Area Multi-Object Fiber Spectroscopic Telescope (LAMOST) \citep{Lamost_cui,Lamost_zhao}, Apache Point Observatory Galactic Evolution Experiment (APOGEE) \citep{APOGEE}, and Galactic Archaeology with HERMES (GALAH) \citep{Galah_DR3,Galah_DR4} provide invaluable amounts of data on elemental abundances, which, coupled with the astrometric and photometric data provided by the $Gaia$ DR3 catalogue \citep{Gaia_dr3}, open up the door to a deeper understanding of our Galaxy.

Concomitantly, recent advances in stellar physics modelling have been made by different research groups, providing numerous stellar evolution models, spanning a wide range of parameters and set of underlying physical assumptions (see \cite{Lebreton_2014_1} for a review). 
Coupled with Bayesian \citep{Jorgensen_Lindegren_2005,Casagrande_2011,Haywood_2013,von_Hippel_2014,SPInS,Cerqui} or probabilistic \citep{Bensby_11} approaches to isochrone fitting, these new datasets and models have yielded millions of stellar ages.

In parallel, missions such as $Kepler$ \citep{Kepler_Borucki,Kepler_Bedding,Kepler_Chaplin}, CoRoT \citep{Corot}, TESS \citep{ricker_14,Hon_2021}, and in the near future PLATO \citep{Rauer_14,goupil_24}, have and will provide data used for asteroseismic age determinations. This method of dating is often complementary to isochrone fitting techniques, as the latter are most often restricted to the main-sequence turn off (MSTO) and sub-giant branch (SGB) of the Hertzsprung-Russell diagrams (HRDs), whereas the former applies to solar-like oscillations of red giant stars, but also main sequence and sub-giants \citep{Lebreton_2014_2,Garcia2019}. Asteroseismic ages have been determined for field stars \citep{Anders_2017,Pinsonneault_2025} as well as clusters \citep{reyes_2025}; however, discrepancies arise when comparing them to ages from isochrone fitting techniques \citep{Tayar_2025}. For a review of the different age determination methods, see \cite{Soderblom_10}.

With the advent of machine learning methods in recent years, age determination techniques have evolved towards more and more data-driven artificial intelligence methods \citep[e.g.][]{Anders_17,Bu_2020,Moya_21,Van_Lane_23,Boulet_24,BNN_stars}. Indeed, machine learning appears well suited for the amount of data made available by large surveys, providing ample training sets. Compared to traditional isochrone fitting methods, which are relatively computationally expensive as they often need to load the full evolutionary grids to run their fitting procedures, neural networks (NNs) have a much lower computational cost. Moreover, once the training procedure has been run and the network has built a faithful representation of the model, the data they have been trained on do not need to be accessed.

Most (but not all; see the discussion in Section~\ref{sec:discussion}) of the machine learning approaches to age determinations are data-driven, meaning they rely on observable quantities, as well as age estimates to be used as a training set. These estimates can be asteroseismic \citep{Das_18,Ciuca_21,Anders_17,Wang_24}, gyrochronologic \citep{Van_Lane_23}, isochrone-fitted \citep{Cantat_gaudin_20}, or chemical-clock \citep{Hayden_22} ages. Depending on the size, coverage, and quality of the training set, this can lead to biases in the resulting predictions, compounded with the biases already present from the methodology that provided these training ages, as well as the biases introduced by the deep learning methodology using this training set. These can be difficult to track down, isolate, and thus deal with. In the best case scenario, if the biases are known, the range of applicability of the deep learning approach can be reduced to a subset where the estimates are reliable. On the contrary, a model-driven deep learning approach would only be dependent on said model and on the methodology to leverage it. In this work, we introduce a model-driven approach to NNs for stellar age determination. Our aim is to apply this technique to a wide range of datasets, and compare our results with other age estimates from the literature. This general approach generates different age distributions, which can also be used to study abundance-age relations, the common structures found between datasets, and their differences.

This paper is structured as follows. In Section~\ref{sec:NN}, we introduce the deep learning methodology, the NN architecture, and the training data and provide a comparison between NNs trained on different stellar evolution models. Section~\ref{sec:data} covers different data samples of cluster and field stars for which the NNs produce age predictions. In Section~\ref{sec:results}, we analyse the age predictions for each sample, as well as age-chemistry relations. In Section~\ref{sec:discussion}, we discuss the advantages and shortcomings of our methodology and its place in the landscape of age estimation methods.

\section{Methods and stellar models}\label{sec:NN}

The general procedure employed is to train an NN on a stellar evolution model grid, taking as inputs observables such as magnitude, colour, and metallicity, and using the age they provide as the output expected from the network. We describe the stellar evolution models used and the architecture of the NN below.

\subsection{Method}\label{sec:NN_architecture}

\begin{figure*}
    \centering
    \includegraphics[width=\linewidth]{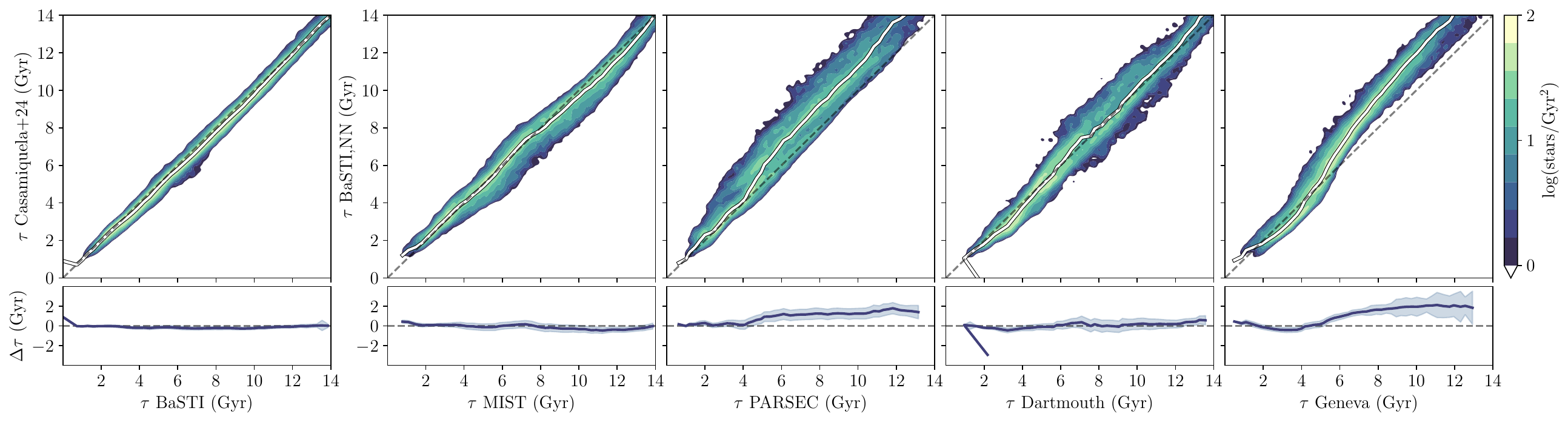}
    \caption{\textit{Top panels, leftmost panel:} Age comparison of \cite{Casamiquela_24}'s 35K sub-giant sample age estimates and results from the NN trained on BaSTI. \textit{Top panels, other panels: } Age estimates from the NN trained on BaSTI compared to those from NNs trained on (from left to right) MIST, PARSEC, Dartmouth, and SYCLIST evolutionary grids. Running medians are shown with white lines, as well as a one-to-one dashed black line to guide the eye. \textit{Bottom panels:} Running median of the difference $\Delta\tau$ between the age estimates $\tau$ of C24 and our BaSTI NN (leftmost panel), between our BaSTI NN and other NNs (other panels) in solid lines, and the associated standard deviations in shaded areas.}
    \label{fig:model_comp}
    \vspace{-0.0cm}
\end{figure*}

We modeled our NNs as multilayer perceptrons, with four hidden layers having 32, 64, 64, and 32 neurons, respectively (see Appendix~\ref{app:architecture} for a representation of the adopted architecture). A lower number of hidden layers as well as a lower number of neurons per hidden layer both lead to decreased performance, as measured by the median absolute deviation (MAD) on the test split, while a higher number of neurons per layer and a higher number of hidden layers both lead to no gain in performance. The networks use a ReLu activation function, and the training was done using the \texttt{MLPRegressor.fit} function of the scikit-learn \citep{scikit-learn} Python package, using the Adam solver \citep{Adam} and a constant learning rate of $1\times10^{-3}$. The number of epochs needed to converge to a constant loss varied depending on the theoretical grid used, with a mean of about 300 epochs.

Similar yet superior results were achieved when the NNs were trained with only three inputs ($\textbf{x}$ = $\{\text{[M/H]},\;M_G,\;(G_{BP}-G_{RP})\}$), as opposed to five inputs ($\textbf{x}$ = $\{\text{[M/H]},\;M_G,\;(G_{BP}-G_{RP}),G_{BP},G_{RP}\}$) (as measured by the MAD on the test split). Thus, we conclude that the addition of separate $G_{BP}$ and $G_{RP}$ magnitudes does not help differentiate between isochrones and that the colour-magnitude space (in our case $M_G,\;(G_{BP}-G_{RP})$) provides most, if not all, of the information needed to infer the age. Different architectures were also tried, with the presented one offering the best compromise between accuracy on the training data and avoiding over-fitting.

Using $T_{\text{eff}}$ as an additional input generally worsened the age estimations. This is at first counter-intuitive as we expect more input data to improve the NN performance. However, in our stellar evolutionary grid, $T_{\text{eff}}$ is strongly correlated to colour, but because our observational data do not have an exact one-to-one relation between the estimated $T_{\text{eff}}$ and colour, our NNs struggle to reconcile observational values of colour and $T_{\text{eff}}$ that do not agree with each other (see Appendix~\ref{app:teff_col} for an example of the theoretical and observed $T_{\text{eff}}-(G_{BP}-G_{RP})$ relation). We note that increasing the number of input parameters also complexifies the NN architecture, which might also impact its ability to converge to a faithful representation of the models. Additionally, $T_{\text{eff}}$ uncertainties are generally high, leading to an age uncertainty being largely dominated by them.

The input parameters $\textbf{x}$ were rescaled before being passed to the NNs by subtracting the mean value and dividing by the standard deviation of each of the training set features. The same rescaling, using the mean and standard deviation of the training set, was performed on the input data passed to the trained NN to estimate ages. The dataset used for training, i.e. the grid points of the stellar evolution models was separated into a training set (90\%) and a testing set (10\%). After training, the network converged to a reasonable representation of the training model, as the MAD on the testing set after the training procedure was about 35 Myr, with variations depending on the grid used.

For a given star, the age estimate $\tau$ is given by the value of the single neuron of the last layer of the NN. This is, however, a single value that does not account for observational uncertainties and, as such, might not provide a good estimate of the true age. Some probabilistic approaches to machine learning, such as Bayesian NNs \citep{BNN}, exist that estimate the uncertainty in their output. However, our training dataset does not include uncertainties, and we are interested in a full age probability description for each star. Ideally, we want to obtain a probability density function (PDF) of the age of the star, given its observable properties $\textbf{x}_{\textbf{obs}}$ and associated uncertainties $\Delta \textbf{x}_\textbf{obs}$. To this end, we followed a Monte Carlo (MC) sampling method, as used in, for example, \cite{Jorgensen_Lindegren_2005}, where we generated a large number of samples $\textbf{x}_\textbf{i} = \mathcal{N}(\textbf{x}_\textbf{obs},\Delta \textbf{x}_{obs})$ (i.e. assuming Gaussian error distributions) and passed them to the NN to obtain age estimates $\tau_i$. The normalised distribution of the age estimates $\tau_i$ thus represents a discretised PDF $p(\tau|x_\text{obs},\Delta x_\text{obs})$, given the observed properties of the star and their uncertainties. From this PDF, we estimated the age uncertainty by computing the standard deviation of the $\tau_i$ age distribution. Different estimators can then be used to obtain a single-valued age from the PDF, namely the median, mean, or mode of the distribution. For a comparison of these three estimators, we refer to \cite{Jorgensen_Lindegren_2005}.

In this work, the median of the distribution was used as our single-valued age estimate, and we used 10 000 MC samples for each star. As we applied the NNs to a variety of large catalogues, covering a wide range of the colour-magnitude diagram (CMD), a choice of 10 000 Monte Carlo samples ensured a faithful representation of the age PDF, even in regions of the CMD with high isochrone degeneracy and overlap. However, in more constrained, well-defined regions, 1000 Monte Carlo samples are sufficient to generate a representative age PDF. In Appendix~\ref{app:MC_samples}, we discuss how the median ages and their associated uncertainties computed in this work did not significantly change when using 1000 Monte Carlo samples. We verified that using the mode or the mean instead did not affect our conclusions: individually, these estimators can lead to different age estimates, but they have little consequence on the analysis of the large datasets used in this work.

\subsection{Stellar evolution models}\label{sec:models}

Neural networks were trained on different stellar evolution models, whose modelling physics can vary; however, none include the effects of stellar rotation on the structure, i.e. all models have $v/v_{crit}=0$. This work uses the following grids of stellar models: Bag of Stellar Tracks and Isochrones (BaSTI)\footnote{\url{http://basti-iac.oa-abruzzo.inaf.it}} \citep{BaSTI}, MESA Isochrones and Stellar Tracks (MIST)\footnote{\url{https://waps.cfa.harvard.edu/MIST}} \citep{MIST}, PAdova and TRieste Stellar Evolution Code (PARSEC)\footnote{\url{https://stev.oapd.inaf.it/PARSEC}} \citep{PARSEC_Bressan,PARSEC_Nguyen,PARSEC_Costa}, Dartmouth Stellar Evolution Program (DSEP)\footnote{\url{https://rcweb.dartmouth.edu/stellar}} \citep{Dartmouth_2007,Dartmouth_2008}, and the Geneva Stellar Evolution Code\footnote{\url{https://www.unige.ch/sciences/astro/evolution/en/database/syclist}} \citep{Geneva_ekstrom,Geneva_georgy_a,Geneva_georgy_b,Geneva_georgy_c}, specifically the large grids of \cite{geneva_yusof}. The parameter spaces for each model are given in Appendix~\ref{app:models}. We chose solar-scaled grids, with non-$\alpha$-element-enhanced abundances, as we corrected our stellar samples for non-zero $\alpha$ element abundances to build a homogeneous set of models. All models include the overshooting of convective cores and microscopic diffusion.

All stellar evolution phases are included in the grids, except pre-MS (pre-Main Sequence). We note that in the rest of this work, this entire set of points has been kept as training data for the NNs, including the MS phase. Neural networks trained only on specific regions of the CMD, restricted to regions where the isochrones are well separated (e.g. MSTO and SGB) have also been tested, but we found no significant gain in their accuracy, as measured by the MAD on the test set (although reducing the training data size and complexity greatly reduces the training time). Thus, we decided on training the NNs on the grids including the MS, so as to retain a more general tool of age estimation. In the MS sequence region, the age estimations come, as expected, with large uncertainties such that a suitable a posteriori cut on the age uncertainty inevitably removes most stars below the MSTO, due to the high density of isochrones in this region.

We investigated the effects that different evolutionary models have on our age estimates. To this end, we trained NNs with the same common architecture, but on each of the grid of models presented in Section~\ref{sec:models}. We used \cite{Casamiquela_24}'s (thereafter C24) sample of 35K sub-giant stars as a control group, with cuts on logg (3.5 < logg < 4.45) and on age uncertainty ($\sigma_\tau/\tau < 10\%$ and $\sigma_\tau < 1$ Gyr) so as to restrict our comparison to regions of low isochrone degeneracy and overlap. The results are summarised in Fig.~\ref{fig:model_comp}. The age estimates in C24 are based on the SPInS tool \citep{SPInS}, which uses a Markov chain Monte Carlo (MCMC) approach on BaSTI isochrones. The ages used in Fig.~\ref{fig:model_comp} are the median ages of the age PDFs it provides.

Because evolutionary grids might use different solar mixtures as reference for their metallicity, it is important to take into account the different zero points adopted so as to make consistent comparisons between grids. To this end, we took the BaSTI grid as a reference metallicity (which uses \cite{Caffau_11}'s solar abundances), computed, and applied the following metallicity offsets. For the MIST grid, which uses \cite{Asplund_2009}'s solar abundances, we applied a -0.057 dex offset; for the Dartmouth grid, which uses \cite{Grevesse_1998}'s solar abundances, we applied a 0.046 dex offset; and for the Geneva grids, which uses \cite{Asplund_2005} solar abundances, we applied a -0.075 dex offset. We did not apply any offset to the PARSEC grid, as it uses the same solar mixture as the BaSTI grid.

The NN trained on the BaSTI grid gives the best agreement with the C24 results out of all the NNs, as it uses the same evolutionary grid (see the leftmost subplot of Fig.~\ref{fig:model_comp}). The only noticeable offset is at very young ages, which have a lower estimate with our method. Such a tight one-to-one relation highlights two results: first, a deep learning approach, trained on theoretical models instead of on other age estimates of samples, can provide at least as accurate stellar ages as other age dating methods. Second, the discrepancies between different age estimation methods applied on the same data is, in a non-negligible part, due to the underlying theoretical evolutionary model used (as seen in the other subplots of Fig.~\ref{fig:model_comp}). This conclusion holds at least for isochrone fitting procedures, whose result heavily depends on the model used. Indeed, our NN-based age estimates, compared to C24 age estimates using the SPInS MCMC approach, only display appreciable dispersion once different models are used and display a very low dispersion otherwise, when using the same input data. That is to say, under the same input data and stellar evolution model, a MCMC method and an NN will converge towards the same age estimates. As the same data and model were used, the low dispersion seen is a result of the different Bayesian and deep learning approaches. The slightly higher dispersion seen for ages $\sim$ 5 Gyr is a result of the metallicity correction we applied, which was not applied in C24. We verified that using uncorrected metallicities yielded better agreement and did not produce the aforementioned feature.

Other age estimates compare differently. Both the MIST and Dartmouth ages show very little offset of the median age versus true age, with median standard deviations of 0.31 and 0.40 Gyr. The final deviation at ages $\tau_\text{MIST}$ > 12.6 Gyr in the MIST ages is due to the age range of the MIST grid, which stops at 12 Gyr (see Appendix~\ref{app:models}), making the prediction of ages older than this unlikely (but possible, as the NN can extrapolate). The PARSEC ages are mostly offset by about 1.5 Gyr. We also note that the distribution of age estimates is not symmetrical around the running median but instead has a longer tail of estimated old ages, with almost no ages being estimated younger by the NN than those found with the BaSTI NN. Finally, the Geneva grid shows younger ages having a low offset, and older ages being lower than BaSTI ones by up to 3 Gyr for the oldest ones.

Notwithstanding training domain restrictions, it is interesting to note that both the PARSEC and Geneva models tend to give systematically lower ages than BaSTI for nearly all ages, mostly incompatible within 1$\sigma$ (see the bottom panels of Fig.~\ref{fig:model_comp}). We find that correcting for different assumed solar mixtures improves the agreement between models, as we compared ages without such corrections and found that for all models, BaSTI ages would be systematically higher than all other models, for nearly all ages.

Comparisons with other models suggest that input physics in stellar models still produces systematic and random uncertainties of the order of 1-2 Gyr in the age determinations. For stars older than about 4 Gyr and younger than 1-2 Gyr, BaSTI ages are systematically older than the PARSEC and Geneva models considered here. MIST and Dartmouth ages show no systematic offset. We emphasise that the differences are not only systematic, with dispersions sometimes very significant at all ages, in particular with the PARSEC and Dartmouth models. For a discussion on stellar modelling shortcomings and their consequences on age determination, we refer to \cite{Howes_2019}. We note that for this comparison, we did not compute the colours from $T_{eff}$, log $g$, and [Fe/H] as could also be done, but instead relied on the colours provided by each model. This can lead to slight differences, as the PARSEC grid, for example, assumes slightly different bolometric corrections than other grids. Comparisons of age estimates using different grids will also depend on the dataset used and its CMD coverage. To highlight this fact, we compare evolutionary models in Appendix~\ref{app:model_comp} as in Fig.~\ref{fig:model_comp}, but using GALAH DR4 as the comparative dataset. In this case, we find better agreement between grids, as stars populate regions of lower isochrone degeneracy and overlap compared to C24's dataset.

To estimate the stochasticity of the training procedure of NNs, we trained ten NNs on the BaSTI grid and estimated ages with each of them using C24's dataset. The resulting median standard deviation is 0.22 Gyr, compared to 0.37 Gyr for the median standard deviation of each star, as computed by the MC sampling method described in Section~\ref{sec:NN_architecture}. Thus, in the vast majority of cases, the age estimate uncertainty budget is dominated by the measurement-derived uncertainties.

\section{Data samples}\label{sec:data}

\subsection{Field stars}

For this study, multiple large-scale spectroscopic catalogues were considered, each providing metallicity and $\alpha$ elements (as defined by each catalogue, and as [Mg/Fe] for GALAH DR4) abundances crossmatched with apparent magnitudes and parallaxes from the $Gaia$ DR3 catalogue \citep{Gaia_dr3}\rev{The following selection criteria were applied}:
\begin{verbatim}
phot_g_mean_flux_over_error > 50,
phot_bp_mean_flux_over_error > 20,
phot_rp_mean_flux_over_error > 20,
parallax_over_error > 10,
ipd_frac_multi_peak < 2,
ipd_gof_harmonic_amplitude < 0.1,
ruwe < 1.1,
parallax > 0.33.
\end{verbatim}
The last condition was added to restrict our samples to stars within a 3 kpc volume around the Sun, as this is the region where we can estimate interstellar extinction with our method. Parallaxes were corrected for zero-point following \cite{Lindegren_21}. The following spectroscopic catalogues were used:
\begin{itemize}
    \item The LAMOST survey Data Release 10\footnote{\url{https://www.lamost.org/dr10}} \citep{Lamost_cui,Lamost_zhao}, containing 7.4 million stars in the Low-Resolution Survey (LRS) and 2.2 million stars in the Medium-Resolution Survey (MRS). After filtering stars excluding those with $\sigma_{\text{[Fe/H]}} > 0.1$ dex and $\sigma_{v_r} > 5$ km s$^{-1}$, and once crossmatched with our Gaia dataset, this survey provides us with [Fe/H] and [$\alpha$/Fe] for 935K stars in the LRS and 88K stars in the MRS.
    \item The GALAH survey, both the Data Release 3 \footnote{\url{https://www.galah-survey.org/dr3/overview}}\citep{GALAH_dr3_De_silva,Galah_DR3} and 4\footnote{\url{https://www.galah-survey.org/dr4/overview}} \citep{GALAH_dr3_De_silva,Galah_DR4}, containing 588K and 917K stars, respectively. Both data releases were used, as they often provide different metallicity estimates for the same star, leading to different age estimates, as well as to highlight the jump in the number of datable stars in the DR4. After applying selection cuts (\texttt{snr\_c3\_iraf > 30}, \texttt{flag\_sp == 0}, \texttt{flag\_fe\_h == 0}, and \texttt{flag\_alpha\_fe == 0} for GALAH DR3 and \texttt{snr\_px\_ccd3 > 30} and  \texttt{flag\_sp == 0} for GALAH DR4), $\sigma_{\text{[Fe/H]}} < 0.1$ dex, and crossmatching with our \textit{Gaia} sample, we end up with 13K and 251K stars with spectroscopic, astrometric and photometric data.
    \item The Apache Point Observatory Galactic Evolution Experiment (APOGEE) Data Release 17 survey\footnote{\url{https://www.sdss4.org/dr17/irspec}} \citep{APOGEE} containing spectroscopic information for 657K stars, of which we recover 147K after crossmatching with our $Gaia$ dataset and filtering out stars with a bad bitmask \texttt{ASPCAPFLAG} flag (bit 23 for \texttt{STAR\_BAD}, bit 19 for \texttt{M\_H\_BAD}, and bit 20 for \texttt{ALPHA\_M\_BAD}), signal-to-noise ratio (S/N) < 50, and $\sigma_{\text{[Fe/H]}} > 0.1$ dex.
\end{itemize}

\begin{figure*}
    \centering
    \includegraphics[width=\linewidth]{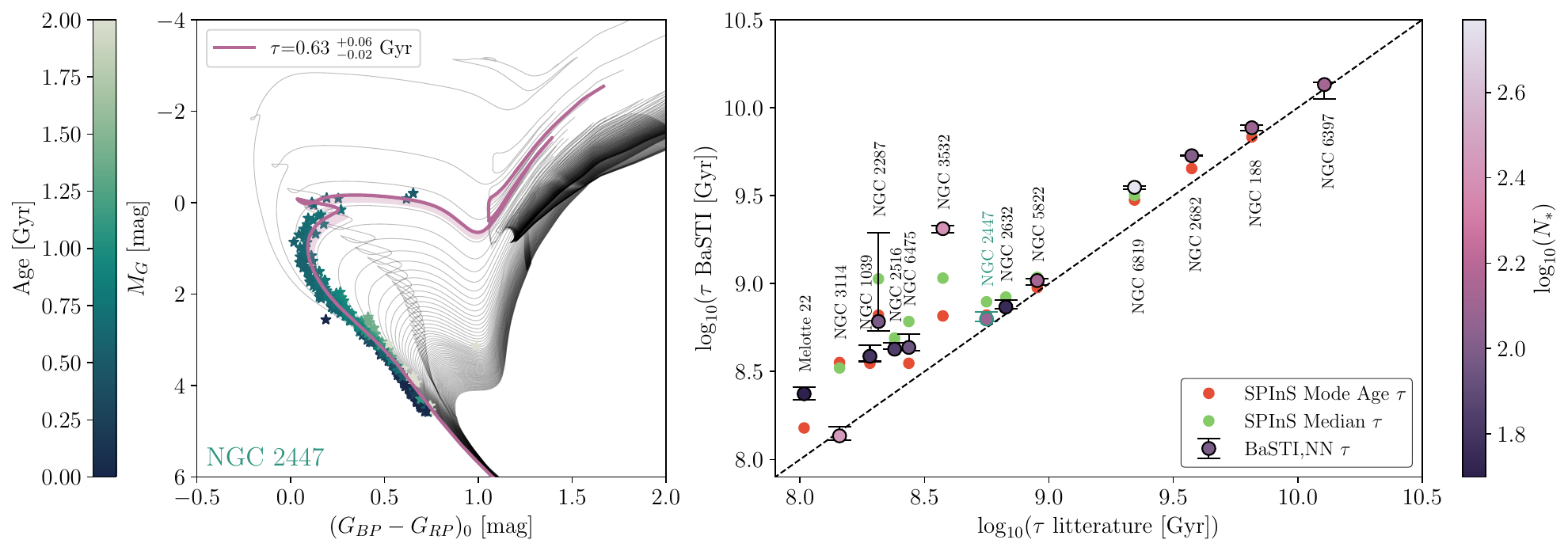}
    \caption{\textit{Left panel:} CMD of cluster NGC 2447 members (star symbols, coloured by our age estimates). Superimposed are BaSTI isochrones in black and the isochrone corresponding to the cluster age given by our method in pink, with its given uncertainties represented by the filled area. \textit{Right panel:} Age comparison of our (pink symbols) and \cite{Casamiquela_24}'s age estimates (mode in red and median in green) and literature ages as defined in C24, coloured by the number of members. NGC 2447 is highlighted in cyan.}
    \label{fig:cluster_comp}
    \vspace{-0.0cm}
\end{figure*}

To correct for interstellar extinction, we used the 3D extinction maps from \cite{vergely}, which cover a 3 kpc x 3 kpc x 0.8 kpc region around the Sun at a 10 pc resolution. We integrated these maps along the line of sight of each star to compute the total $A_V$ extinction value. Because extinction both dims and reddens light, it is vital to implement it as faithfully to the ground truth as possible, as any deviation will inevitably affect age estimation. For a discussion of these effects and a comparison with \cite{edenhofer} extinction maps, see Appendix~\ref{app:av_comp}. The V-Band $A_V$ values computed were translated into $Gaia$ passband interstellar extinction values $A_G,A_{BP}$, and $A_{RP}$ using the relations given in \cite{Danielski}, and $Gaia$ eDR3 coefficients\footnote{\url{https://www.cosmos.esa.int/web/gaia/edr3-extinction-law}}. Outside the range of applicability for this relation, we used \cite{Wang_2019}'s relation. The CMDs and galactic coverages of each survey are presented in Appendix \ref{app:surveys}.

As the grids used in this work all have solar $\alpha$ element abundances, the global metallicity $\text{[M/H]}$ was corrected to take the non-zero $\alpha$ element abundances of stars into account by applying the \cite{Salaris} relation, with coefficients corresponding to the \cite{Caffau_11} solar abundances used by the BaSTI grid:
\begin{equation*}
    \text{[M/H]} = \text{[Fe/H]} + 0.76{[\alpha/\text{Fe}]}.
\end{equation*}

For a comparison of ages with and without this rescaling step to study the influence of $\alpha$ elements on age determination, see Appendix~\ref{app:alpha_comp}. We note that we did not correct for different solar mixtures used by the spectroscopic catalogues, as we did for the different evolutionary grids in Section~\ref{sec:models}. This is motivated by the fact that most age catalogues we compare our results to do not apply these corrections -- even those using stellar evolutionary models that do not share the same solar mixtures as the spectroscopic surveys they use. Moreover, the LAMOST survey is based on a data-driven training approach, making a metallicity correction more difficult. Thus, we decided not to apply these corrections and instead made comparisons with metallicities as is.

The final uncertainties on the extinction-corrected $M_G$ and $(G_{BP}-G_{RP})_0$ were obtained by propagating the uncertainties on the apparent magnitudes in the $G$, $BP$, and $RP$ bands, the uncertainty of the $Gaia$ parallaxes, and the uncertainty of the coefficients of the relations of \cite{Danielski} and \cite{Wang_2019}. We note that these are lower bounds as no uncertainty on $A_V$ is provided by the interstellar extinction maps. This set of uncertainties, along with $\sigma_\text{[M/H]}$, forms the vector $\Delta \textbf{x}_\textbf{obs}$ used to sample Monte Carlo realisations around the mean position $\textbf{x}_\textbf{obs}$ of each star, as described in Section~\ref{sec:NN_architecture}.

\subsection{Stellar clusters}
In addition to field stars, we also estimate ages of open and globular clusters and compare them to other age estimates in the literature. To this end, we take the selection of clusters given in C24, which contains 13 open clusters and one globular cluster. We assume the same extinction values, metallicities, and distances, as defined in their work (see their Table 1). The literature ages are taken from \cite{Gaia_18,Cantat_gaudin_20,Netopil_22,Tsantaki_23}.

\section{Results}\label{sec:results}

\subsection{Age estimation of stellar clusters}\label{sec:cluster_result}

Because clusters are usually thought of as coeval stellar populations, they offer a wide coverage in the CMD of what is considered as a single age isochrone sequence. Most isochrone fitting methods benefit from this fact and use each star member as a unique data point along a shared sequence to constrain the cluster age, although it is worth noting that this deceivingly simple assumption does not hold for some globular clusters, which may contain multiple stellar populations of different ages and chemical compositions (see \cite{Bastian_18} for a review). Our method, in contrast, only considers, by construction, single independent stars when dating clusters, as our NNs have been built to receive single star inputs. While this is useful if one is interested in investigating separate populations within a cluster, we need a way to combine the individual age estimates produced to form a cluster age estimate.

Mathematically speaking, the global cluster age PDF is defined as the product of the individual star age PDFs. However, in practice, not all PDFs have a non-zero contribution at the true cluster age, such that a few individual PDFs can make the global product of PDFs vanish. This problem was mentioned in \cite{Jorgensen_Lindegren_2005}, who solved it by adding a small $\varepsilon$ contribution to all PDFs so that their product did not involve multiplication by zero:

\begin{equation}\label{eq:pdf}
    p_\text{Cluster}(\tau|\textbf{x}_\textbf{obs}) = \prod_{n=1}^{N_\star}\left(p_i(\tau_{\text{}}|\textbf{x}_\textbf{obs,i})+\varepsilon\right).
\end{equation}

In some pathological cases, if the cluster CMD sequence is broad enough or has enough outliers (e.g. binaries or blue stragglers), the computed PDF can still vanish. To remedy this, we employed a bootstrapping approach, where we computed $p_\text{Cluster,j}$ from a random sample of our $N_{\star}$ cluster members, with replacement. This step was iterated a number of times (in our case 100) to produce a set of $p_\text{Cluster,j}$ containing some non-zero PDFs when enough outliers had been discarded. The cluster age was then taken to be the median of the peaks of these well-defined $p_\text{Cluster,j}$.

The age PDFs of each bootstrap step generally peak at a single value, such that $p_\text{Cluster,j} \sim \delta(\tau-\tau_\text{Cluster,j})$, providing no information on the global uncertainty of this estimate. Thus, we estimated the age uncertainty of the cluster by computing the standard deviation of the distribution of the peaks of our $p_\text{Cluster,j}$. This value, providing an estimate of the random variations in the cluster age given a subsample of its members, is usually very small and should not be interpreted as a total error estimate. In fact, the biggest source of error would be systematic, from the physics of the chosen evolutionary grid, the extinction and distance computations, as well as metallicity and $\alpha$ elements abundance estimations, all of which were not included in this random error estimate.

The results of the cluster age estimation are presented in Fig.~\ref{fig:cluster_comp}. In the left panel, we give an example of a cluster (NGC 2447) sequence in a CMD, where each cluster member is coloured by its estimated age. We over-plot the isochrone age at $\tau$=0.63 Gyr. This age is very close to other ages given for this cluster in the literature (e.g. \cite{Cantat_gaudin_20} give an age of 0.575 Gyr; \cite{Hunt_23} an age of 0.74 Gyr; \cite{Cavallo_24} an age of 0.63 Gyr; and \cite{Bossini_19} an age of 0.56 Gyr). Indeed, the isochrone closely follows the cluster sequence, from the MS to the MSTO. 

In the right panel of Fig.~\ref{fig:cluster_comp}, our age estimates are compared to those of C24, using their mode (red) and median (green) age estimates, and to ages from the literature, as defined in C24. Overall, we find good agreement between our ages and literature ages, particularly for clusters with $\tau \gtrsim$ 1 Gyr. For younger clusters, we find the same trends as in C24, i.e. our ages are higher than those found in the literature, for which the relative difference of estimates can be higher. We also find that this effect can be explained by stars in the MS, which tend to have overestimated ages compared to the cluster age and MSTO stars. The former occupy a more isochrone-dense region of the CMD, where age estimates produce more uniform PDFs during the Monte Carlo sampling process, leading to higher individual ages.

NGC 2287 has a higher uncertainty than other clusters, as its CMD sequence does not follow a single isochrone. In the MS, its members overlap with isochrones of age 1 to 1.5 Gyr, while in the MSTO, its members more closely follow isochrones of age $\sim$ 0.4 Gyr. The literature age is in fact $\sim$200 Myr, much younger than our age estimates for MS members, but closer to our age estimate for its MSTO members. A similar situation is found in NGC 3532, where the MS aligns with isochrones of $\sim$ 2 Gyr, whereas MSTO align with isochrones of age < 0.5 Gyr, closer to the literature age of $\sim$ 350 Myr. In this case, the low uncertainty is driven by two probably spurious stars on the $\tau=14$ Gyr isochrone, far from the cluster sequence, driving the computed $p_\text{Cluster}(\tau|\textbf{x}_\textbf{obs})$ to a narrow peak at 2 Gyr. This discrepancy can also be seen in C24 (Fig. 5), where both clusters have a long tail in their PDF towards higher ages.

For clusters of ages $\log_{10}(\tau[$Gyr$])$ > 9, our age estimates agree with both C24's and literature results, even if they are still all larger. On all clusters, we find a MAD between our estimates and the mean literature age of 0.20 Gyr, compared to a MAD of 0.27 Gyr for C24's results with SPInS.

\subsection{Age estimation of field stars}\label{sec:age_comp}

\begin{figure*}[!htbp]
    \centering
    \includegraphics[width=\linewidth]{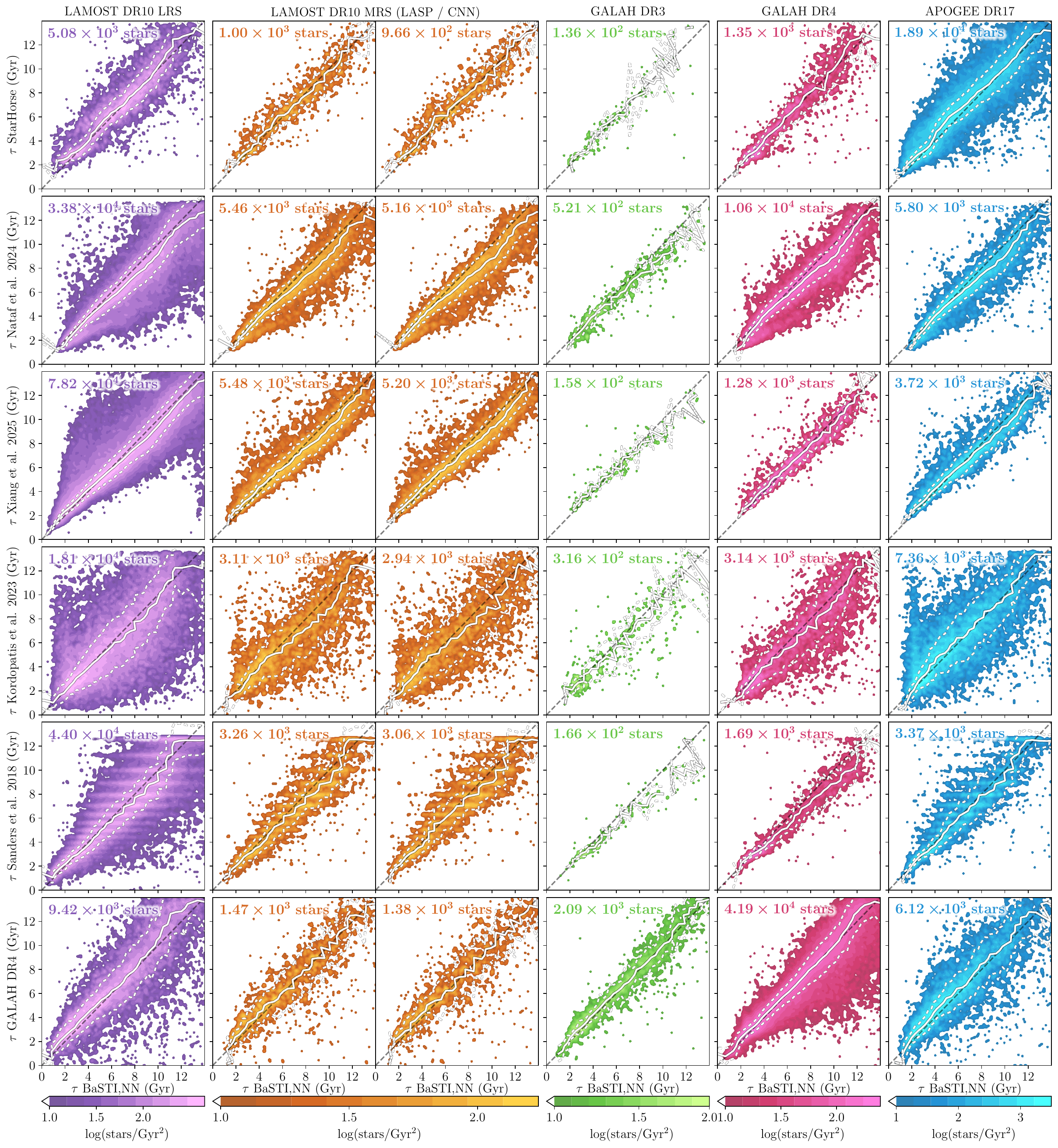}
    \caption{Age comparison of the ‘golden’ samples as defined in Section~\ref{sec:age_comp} used in this work with age estimates from the literature. From top to bottom: StarHorse \citep{Starhorse_catalog}, \cite{Nataf}, \cite{Xiang_2025}, \cite{Kordopatis}, and \cite{Sanders}. The number of stars is shown in each panel in the top left corner, a one-to-one dashed line is provided to guide the eye. The running median and $\pm1\sigma$ are plotted in solid white and dashed lines, respectively.}
    \label{fig:age_comparison_literature}
    \vspace{-0.0cm}
\end{figure*}

We then applied our NNs to the data samples of field stars described in Section~\ref{sec:data}. This \rev{is} a more challenging task than dating clusters, as field stars are not part of a coeval population but rather individual, unrelated points in the CMD. Nevertheless, the NN provide an age value, whose precision can then be appreciated through the uncertainty estimated by the method described in Section~\ref{sec:NN_architecture}. Henceforth, we restricted each sample of ages obtained to ‘golden’ samples by imposing the following criteria:
\begin{itemize}
    \item The star's position in the CMD, within its uncertainties ($\sigma_{M_G}$, $\sigma_{(G_{BP}-G_{RP})_0}$,$\sigma_\text{[M/H]}$), should lie within the training domain of the NN, i.e. the set of evolutionary curves of the model used for training\footnote{The training domain is computed by building a grid where each cell is populated by stellar tracks passing through it. Convex hulls are not suitable, as tracks have concave shapes, and we experimented with $\alpha$ shapes but found the discretized representation gave better results in thin areas such as the MS.}.
    \item The relative age uncertainty $\frac{\sigma_\tau}{\tau}$ should be less than 10\%.
    \item The absolute age uncertainty $\sigma_\tau$ should be less than 1 Gyr.
    \item The age estimate $\tau$ should be between 0 and 15 Gyr.
    \item The star's surface gravity log $g$ should be between 3.5 and 4.45. This criterion is enforced to restrict our comparison to stars for which we expect the isochrones to be separated and reliable enough for dating stars. On the lower end, we remove red giant stars, for which we know general isochrone fitting methods struggle, and on the higher end, we removed dwarfs in the MS where isochrone density was too high to provide a reliable age information. The bounds were set manually to minimise contamination in our samples and maximise the number of remaining stars.
\end{itemize}

For each NN trained on a different evolutionary model, we thus obtained an age catalogue for our stellar samples. In this section, we compare the ages we obtained with the NN trained on the BaSTI grids to other age estimates from the literature:
\begin{itemize}
    \item The StarHorse catalogue \citep{Starhorse_catalog}, which uses stellar photometry and PARSEC isochrones to estimate ages based on a Bayesian approach. It provides ages for stars in the APOGEE, $Gaia$-ESO, GALAH, and RAVE catalogues, for a total of more than 10 million stars.

    \item The \cite{Nataf} (\textbf{N24}) catalogue of 401K stars dated through isochrone fitting using the MIST isochrone set.
    \item The \cite{Xiang_2025} (\textbf{X25}) catalogue of 320K stars, which extends the \cite{Xiang_2022} catalogue of 247K stars, also dated through an isochrone fitting procedure, using the Y$^2$ isochrone \citep{YY_Yi,YY_Demarque} set.
    \item The \cite{Kordopatis} (\textbf{K23}) catalogue of $\sim$ 5.5 million stars dated through isochrone fitting using the PARSEC and COLIBRI \citep{COLIBRI_Marigo,COLIBRI_Rosenfield,COLIBRI_Pastorelli} sets of isochrones.
    \item The \cite{Sanders} (\textbf{S18}) catalogue of 3.7 million stars from the APOGEE, $Gaia$-ESO, GALAH, LAMOST, RAVE, and SEGUE spectroscopic surveys, through a Bayesian approach and PARSEC isochrones.
    \item The GALAH DR4 \citep{Galah_DR4} (\textbf{GDR4}) catalogue of stellar ages, providing ages for 900K stars through a Bayesian approach and PARSEC isochrones.
\end{itemize}

The results are shown in Fig.~\ref{fig:age_comparison_literature}, where each column represents one of our data samples and each row is one of the age catalogues used for the age comparison. The number of stars crossmatched between each sample and each catalogue are displayed in each panel. The running median is shown as a white line and was computed as med($\tilde{x}=\frac{x-y}{\sqrt{2}}$) in diagonal bins $\tilde{y}_i<\tilde{y}\leq\tilde{y}_{i+1}$ with $\tilde{y}=\frac{x+y}{\sqrt{2}}$. The same was done for the 16th and 84th percentiles, shown in dashed white lines. Indeed, computing the running median in bins of x or y instead can result in boundary issues. For example, at 13 Gyr < x $\leq$ 14 Gyr, the \rev{age distribution along the y axis is} not symmetrical, with most ages between 13-14 Gyr -- some below but very few above. Thus, the median is biased towards younger ages; similarly, but in the opposite direction, for running medians computed in y bins. Computing the median along the diagonal direction instead ensures the result is independent of the bins chosen to compute it and unbiased close to the boundaries.

Overall, our method provides age estimates in good agreement with ages from the literature. \textbf{K23} ages, while being in good agreement in the most dense regions for all of our samples shows the highest standard deviation, with a median standard deviation of 1.61 Gyr marginalised on all our samples. \textbf{GDR4}'s age estimates also show a relatively high standard deviation on some samples with a median standard deviation of 1.21 Gyr, marginalised on all ages and samples. For \textbf{S18}, there are clear differences at the high end of the age range, with an absence of any ages $\tau > 13$ Gyr. \textbf{X25} shows the lowest standard deviation, with a median value of 0.79 Gyr. All of our catalogues show good agreement with our results. This is especially true for the \textbf{X25} catalogue, which displays a very low to zero offset and the lowest standard deviation. LAMOST DR10 MRS (both LAMOST Stellar Pipeline (LASP) and convolutional neural network (CNN)), consistently display the highest dispersion across all of the samples and the greatest divergence from the one-to-one relation -- with age estimates for the oldest ages higher by a few gigayears than most catalogues. For \textbf{X25} and \textbf{N24}, the median diverges at ages > 5 Gyr. While for other catalogues, the median has very little offset from the one-to-one relation. Overall, the fact that the age catalogues and the different samples generally agree and have a low systematic difference is surprising, as one would expect the variety of methods, evolutionary tracks, extinction and distance estimates used to have a strong effects on the age estimates.

\begin{figure*}
    \centering
    \includegraphics[width=.75\linewidth]{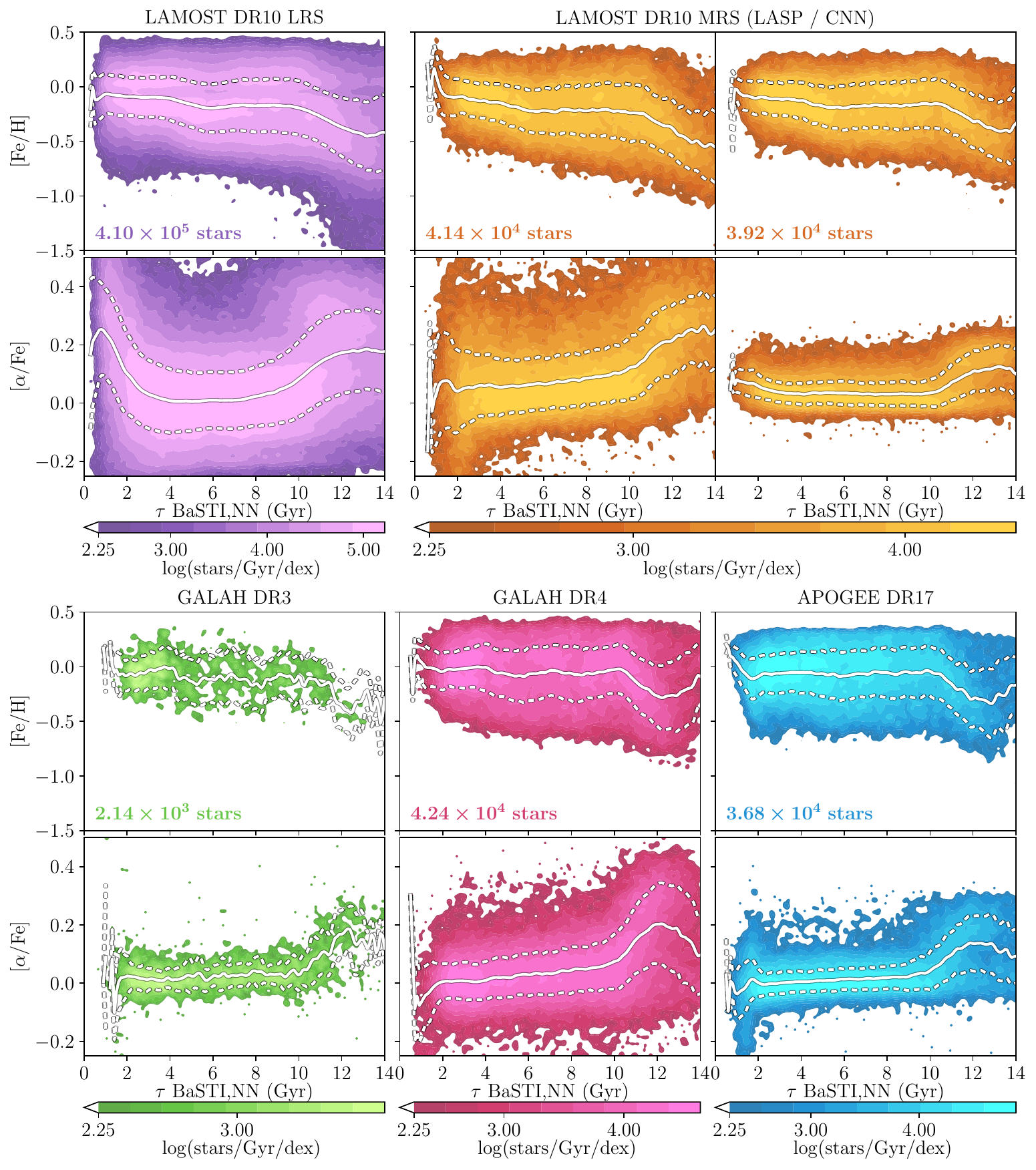}
    \caption{Abundance vs age planes for our samples, with ages estimated with the NN trained on the BaSTI grid. \textit{Odd rows:} Age vs [M/H]. \textit{Even rows:} Age vs [$\alpha$/Fe]. The number of stars in each sample is annotated in the corners of the upper panels. The lighter colours correspond to higher density regions, and the means and $\pm1\sigma$ of the distributions are plotted as solid white and dashed lines, respectively.}
    \label{fig:age_abundance}
    \vspace{-0.0cm}
\end{figure*}

\subsection{Age-chemistry relations}\label{sec:abund}

Using the golden samples defined in the previous section and the spectroscopic abundances, we plot the age-abundance diagrams in Fig.~\ref{fig:age_abundance}, with age versus metallicity in the top row and age versus $\alpha$ element abundance in the lower row. Interestingly, the age-metallicity distributions do not share the same patterns. Some, such as LAMOST DR10 MRS (LASP and CNN) and GDR4, display two sequences at ages $\tau \lesssim 10$ Gyr, commonly seen in other studies of our Galaxy's age-metallicity relation (AMR) \citep{Haywood_2013,Bensby_2014,Xiang_2022,Cerqui}. For ages $\tau \gtrsim$ 10 Gyr, they show a sharp decline compared to the flat metallicity for younger ages, more prominently shown in LAMOST DR10 MRS (LASP and CNN) and GDR4, seen albeit more faintly in LAMOST DR10 LRS. We note that the under-density at all ages at [Fe/H] $\sim$ -0.1 for LAMOST DR10 LRS is not a result of our selection cuts but seems to be inherent to this specific data release of the LAMOST survey. The distribution for GALAH DR3 is more noisy on account of the lower number of stars; however, the two sequences are still visible.

Similarly, the age-$\alpha$ distributions, while sharing similar features, vary from sample to sample. In all our samples, a clear thick-disk sequence can be seen at ages $\tau \gtrsim$ 10 Gyr, with a sharp increase in [$\alpha$/Fe] from $\sim$ 0 up to 0.4 -- particularly for a number of stars in the GDR4 sample, which show the sharpest increase in $\alpha$ element abundance. We note a similar over-density of $\alpha$-rich stars (> 0.1) in LAMOST DR10 LRS, MRS LASP, and APOGEE DR17 for stars with ages 0 Gyr $\lesssim \tau \lesssim 4$ Gyr. We note that the definition of the $\alpha$-element content, which we took directly from the catalogues and as [Mg/Fe] for GDR4, varies from survey to survey, explaining the stark differences in [$\alpha$/Fe] distributions (e.g. between APOGEE DR17 and GDR4 or LAMOST DR10 LRS). Nevertheless, the difference between LAMOST DR10 MRS LASP and LAMOST DR10 MRS CNN is striking (as also seen in C24), considering the two share the same set of observed stars and differ only in their metallicity estimates.

Imposing stricter cuts on the age uncertainty by for example filtering out stars with $\sigma_\tau$ > 0.3 Gyr reveals clearer sequences for LAMOST DR10 (LRS and MRS), as well as for APOGEE DR17. The number of stars with solar-value $\alpha$ element abundances with ages $\tau \gtrsim$ 10 Gyr is greatly reduced, making the sequences both in the [Fe/H]-age and the [$\alpha$/Fe]-age relations thinner. However, as the age error generally increases with age, a strict criterion on absolute rather than relative age uncertainty greatly limits the analysis of older populations. For example, when imposing the $\sigma_\tau$ > 0.3 Gyr criterion, we lose most stars with $\tau$ > 10 Gyr in our GALAH DR3 and DR4 sample.

\begin{figure*}[!htbp]
    \centering
    \includegraphics[width=\linewidth]{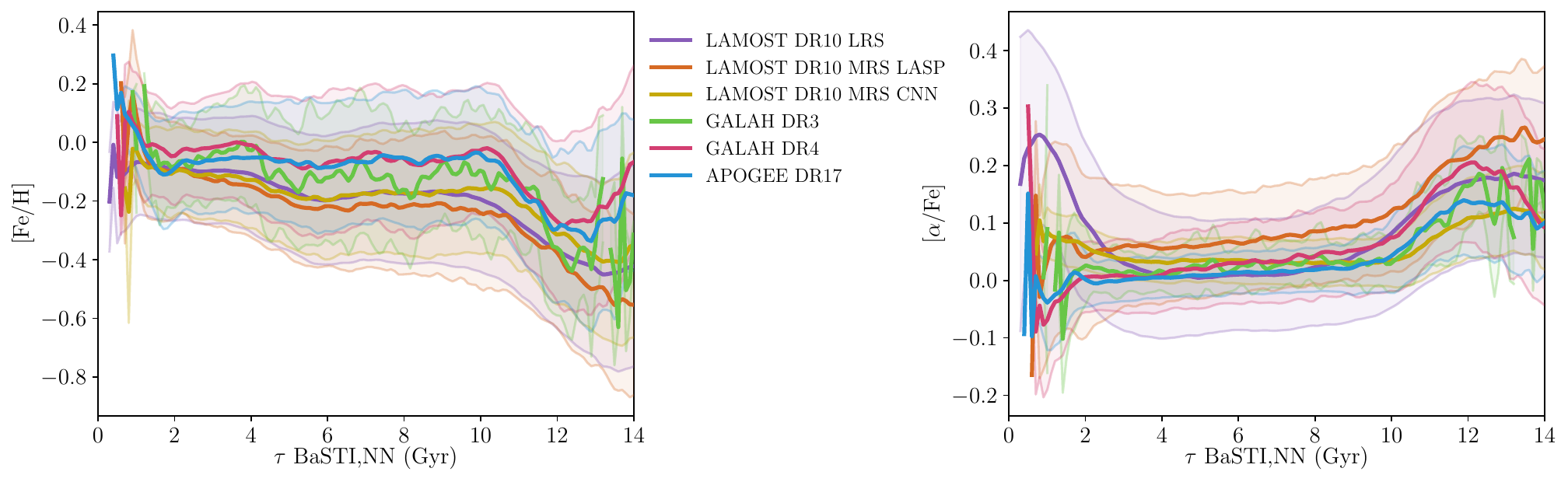}
    \caption{Abundance vs age trends for each of our samples, with ages estimated with the NN trained on the BaSTI grid. The coloured lines and shaded areas represent the means and $\pm1\sigma$ of the distributions, as shown in white in Fig.~\ref{fig:age_abundance}.}
    \label{fig:age_abundance_trends}
    \vspace{-0.0cm}
\end{figure*}

The distributions in the age-abundance planes of our six samples seem to vary drastically, for example, between the broad LAMOST DR10 LRS and the thin LAMOST DR10 MRS CNN distribution. However, taking the mean [Fe/H] and [$\alpha$/Fe] for each age bin reveals similar trends. In Fig.~\ref{fig:age_abundance_trends}, we plot the mean [Fe/H] and [$\alpha$/Fe] of each of our samples, with age estimates from the NN trained on the BaSTI grid. We find good agreement between the different samples, both in the AMR and in the [$\alpha$/Fe]-age relation. At young ages $\tau$ < 1 Gyr, the samples diverge from each other, mostly due to the very low number of stars. For ages $\tau$ > 1 Gyr, the mean relations closely follow each other, with a flat to shallow sequence from 1 to 10 Gyr in the AMR, with an initial metallicity of -0.1 for all samples at 2 Gyr, and a final metallicity between -0.2 and -0.1 at 10 Gyr. For older ages $\tau$ > 10 Gyr, the AMRs show a steeper slope, reaching metallicities between -0.2 and -0.5 at 14 Gyr. For GALAH DR3 and DR4, metallicity drops for ages between 5 and 9 Gyr; this can also be seen -- albeit faintly -- in the LAMOST DR10 samples. The six AMRs share the same knee at 10 Gyr, separating the two sequences. The same behaviour is seen in the [$\alpha$/Fe]-age relations, with an initial divergence of the samples at young ages $\tau$ < 1 Gyr, followed by a shallow to flat sequence between 1 and 10 Gyr. All samples except LAMOST DR10 MRS LASP follow a very tight sequence centred at [$\alpha$/Fe] = 0.025, while LAMOST DR10 MRS LASP is offset by $\sim$ 0.05 dex. At older ages $\tau$ > 10 Gyr, the relations show steeper slopes, reaching [$\alpha$/Fe] abundances between 0.1 and 0.2. For a comparison of the [Fe/H]-age and [$\alpha$/Fe]-age relations from our APOGEE DR17 sample (with ages estimated by our NNs trained on different models), see Appendix~\ref{app:age_abund}.

\section{Discussion}\label{sec:discussion}

Our approach is one of many to provide stellar age estimates. Compared to traditional isochrone fitting procedures and Bayesian methods, ours is generally much faster, owing to the efficient representation of evolutionary models by a small number of neurons. The advantage of building an embedding space of the model, rather than relying on the model itself, is twofold: first, this leads to fewer overall operations to estimate ages from observational inputs; second the NNs were trained to build this embedding, and as such the initial models are not needed for age determination once the networks are trained. This results in significantly faster estimations, even when accounting for the N Monte Carlo needed to compute the related uncertainties. Compared to SPInS \citep{SPInS}, which takes about 20 seconds to calculate the age of a single star, our approach can estimate the ages of 60,000 stars in the same amount of time (together with 1,000 Monte Carlo samples for each star). We stress that the comparison made is not one-to-one as SPInS provides full PDFs for each of the inferred parameters. However, if one is only interested in an age estimate, for which our method gives similar results, deep learning techniques have a clear advantage. In addition, the computation time is dependent on the HRD region in SPInS, as the convergence time of the MCMC walkers towards a final age PDF varies. This runs contrary to our deep learning approach, which has a constant number of numerical calculations, independently of the inputs and uncertainties. To our knowledge, this is the first time a general age comparison has been made between different stellar evolutionary models, mainly thanks to the speed of our deep learning approach.

As we demonstrated in Section~\ref{sec:models} and in the first panel of Fig.~\ref{fig:model_comp}, our models are in fact able to build a faithful embedding of evolutionary grids. By construction, the NNs are able to interpolate between the learned tracks, which, in traditional isochrone fitting procedures, is an additional non-trivial task \citep{Jorgensen_Lindegren_2005,SPInS,Aguirre_22} for which several approaches are possible. In this work, out of all fundamental stellar parameters, we focused solely on age, contrary to other methods based on evolutionary grids, which can also provide estimates for stellar mass and radius, as well as other properties such as metallicity, extinction, and distance. As this study aims to show deep learning's potential for handling ever-larger datasets quickly and reliably, we focused on age, as it is arguably the most important fundamental parameter in Galactic archaeology. However, as mass and radius are also provided by evolutionary tracks, one could extend our approach to build an NN trained to predict those parameters as well, with a minor impact on computational time.

Due to our choice of extinction maps, the age estimates of this work are confined to stars within a 3 kpc cylindrical radius around the Sun, which is the volume of the high-resolution maps of \cite{vergely}. Our analysis could be extended to higher distances from the Sun by using the lower-resolution maps of, for example, \cite{vergely}, or \cite{edenhofer}. We stress that the results would vary, as ages depend on $A_V$ estimates (see Appendix~\ref{app:av_comp}); however, the conclusions should remain similar as we did not observe any overall offset between age estimates using different extinction maps.

Upcoming surveys such as WHT Enhanced Area Velocity Explorer (WEAVE) \citep{Jin_2024}, Sloan Digital Sky Survey-V (SDSS-V) \citep{Kollmeier_2025}, and 4-metre Multi-Object Spectroscopic Telescope (4MOST) \citep{4MOST} will provide an-order-of-magnitude increase in stellar element abundances, enabling age estimates and more comprehensive studies of our Galaxy's history and evolution. A deep learning approach will provide fast and reliable age estimates of the large datasets these surveys will deliver.

Other catalogues, such as \cite{Zhang_catalog,Anders_22,Andrae_23,Khalatyan_24,Martin_24,Guiglion_24,Li_24}, provide stellar atmospheric parameters for a very large number of stars by leveraging machine learning methods for their estimation \citep{Anders_23}. As an example, \cite{Zhang_catalog} trained their model on data from the LAMOST survey to model $Gaia$ XP spectra, which they then applied to the 220 million available GDR3 XP spectra to obtain estimations of their stellar atmospheric parameters. Notably, their catalogue does not contain [$\alpha$/Fe] estimates. However, as shown in Appendix~\ref{app:alpha_comp}, $\alpha$ element abundances can have a non-negligible effect on age estimations, especially for older stars for which \cite{Salaris}'s rescaling relations of metallicity can induce a shift of several 0.1 dex. For younger stars, whose $\alpha$ element contents are close to solar, we expect our method to provide reliable age estimates using [Fe/H] rather than the rescaled [M/H]. As such, open clusters, which contain young stellar populations, can benefit from catalogues such as \cite{Zhang_catalog}, which provide broad isochrone sequences at well-known metallicities.

Finally, we highlight the work of \cite{Garraffo_21}, which also developed a model-driven deep-learning age-determination method based on MIST evolutionary tracks. However, theirs was based on the $T_\text{eff}$-luminosity-age relation and was only applied to solar-metallicity stars as it did not account for metal content. \cite{Hon_24} also trained generative models on MESA stellar tracks \citep{Paxton_11,Paxton_13,Paxton_15,Paxton_18,Paxton_19}. They used asteroseismic observables rather than photometric data as input, aiming to produce an emulator for evolutionary tracks.

The NNs developed for this work are made available through \texttt{NEST}, a Python package, as well as through a web interface. The \texttt{NEST} package includes individual age estimations computed using the method described in Section~\ref{sec:NN_architecture}, as well as cluster age estimations using the method described in Section~\ref{sec:cluster_result}. The left panel of Fig.~\ref{fig:cluster_comp} was plotted using its \texttt{HR\_diagram()} method, plotting a set of isochrones as well as an interpolated isochrone of the cluster age.

\section{Conclusions}

In this paper, we introduced a model-driven deep-learning approach to stellar age determination. We trained deep NNs on stellar evolution models using $Gaia$ photometry and metal and $\alpha$ element abundances to predict ages.

We trained multilayer perceptrons on different stellar grids (BaSTI, MIST, PARSEC, Dartmouth, and Geneva) until their loss function converged. The resulting MAD on the test data is $\sim$ 35 Myr, confirming the faithful representation of the model by the network. Following \cite{Jorgensen_Lindegren_2005}, we used Monte Carlo sampling to estimate uncertainties on our age estimates.

Contrary to traditional isochrone fitting methods, ours is generally much faster, owing to the low computational cost of NNs, with an estimated 60,000 speed-up compared to an MCMC procedure used in, for example, the SPInS pipeline. Other deep learning approaches, which are generally data-driven (i.e. trained on a dated subset of stars), typically suffer from biases inherited from the training sample's age estimation method and sample size and can be restricted to a subset of the available star samples. In comparison, our method only depends on the evolutionary tracks used and offers a broader range of application.

We compared age estimates made with NNs trained on different evolutionary tracks and find some disagreements between tracks, with PARSEC tracks yielding lower ages than BaSTI by $\sim 1-2$ Gyr, but relatively low overall dispersion. Geneva tracks show a much higher divergence, with old ages being lower than SPInS ones by as much as 4 Gyr. When comparing our method using BaSTI tracks with the results of C24 using the SPInS pipeline \citep{SPInS} and BaSTI tracks, we find no offset for all ages except at very low ages, confirming the reliability of our method. Thus, we find that differences in age estimates appear mostly from using different evolutionary tracks and/or from using different input data, such as extinction estimates (see Appendix~\ref{app:av_comp}).

We applied our NNs to stellar cluster populations, from which we derived a global cluster age by combining the individual age PDFs. We find that our cluster age estimates are in agreement with the literature for older ages, while our ages are systematically higher than the literature age for younger clusters. This effect was also seen in the analysis of \cite{Casamiquela_24}, who also used the BaSTI grid. We find a MAD between our ages and literature values of 0.20 Gyr, compared to 0.27 Gyr for \cite{Casamiquela_24} using SPInS.

We then applied our NNs to large spectroscopic catalogues, namely LAMOST DR10, GALAH DR3, GDR4, and APOGEE DR17, crossmatched with $Gaia$ photometry and parallaxes. The photometry was corrected for extinction using \cite{vergely} 3D extinction maps. The resulting ages were compared to the other age catalogues of StarHorse \citep{Starhorse_catalog}, \cite{Nataf} (N24), \cite{Xiang_2025} (X25), \cite{Kordopatis} (K23), \cite{Sanders} (S18), and \cite{Galah_DR4} (GDR4). We find overall good agreement, especially for the X25, N24, and StarHorse catalogues, where the running median of our age versus theirs shows low to no offset and low dispersion. K23 and GDR4 both show a relatively low offset, giving lower ages compared to ours for older ages, as well as higher median dispersions of 1.61 and 1.21 Gyr, respectively. The X25 catalogue shows the least dispersion, with a median value of 0.79 Gyr \rev{when comparing our ages and theirs}, and very low differences in median age estimates. We find that LAMOST DR10 MRS displays the highest dispersion for all catalogues. Aside from divergences at older ages in the X25 and N24 catalogues, the running median closely follows the one-to-one relation for all other catalogues and samples.

Finally, we analysed the abundance-age relations of the aforementioned spectroscopic catalogues and find similar structures, with a thin and thick-disk sequence visible both in the [Fe/H]-age and [$\alpha$/Fe]-age planes. For stars younger than $\sim$ 10 Gyr, we find a flat metallicity and $\alpha$ element distribution with age, and for older stars, we find a sharp decrease in metallicity and a sharp increase in [$\alpha$/Fe]. The epoch separating these two sequences is also noticeable when comparing ages estimated using metallicity scaling relations based on $\alpha$ element content with those obtained without such scaling relations (see Appendix~\ref{app:alpha_comp}). Given its flexibility and speed, this code opens unprecedented possibilities for estimating ages of millions of stars in upcoming spectroscopic surveys such as 4MOST \citep{4MOST}, SDSS-V \citep{Kollmeier_2025}, and possible future ones such as WST \citep{WST}.

Parallel to this work, the trained NNs are released as part of \texttt{NEST}\footnote{\url{https://github.com/star-age/NEST}}, a Python package to infer individual and population ages and estimate uncertainties. We also release a web interface\footnote{\url{https://star-age.github.io}} that incorporates the same trained NNs and an interactive CMD for dating stars, as well as tools for importing star populations to be dated.

\section*{Data availability}
Full catalogues are available at the CDS via anonymous ftp to \href{https://cdsarc.cds.unistra.fr}{cdsarc.cds.unistra.fr} \href{ftp://130.79.128.5}{(130.79.128.5)}
 or via \url{https://cdsarc.cds.unistra.fr/viz-bin/cat/J/A+A/000/A000}. The \texttt{NEST} python package is available at \url{https://github.com/star-age/NEST} and the web interface at \url{https://star-age.github.io}.

\begin{acknowledgements}
T.B. thanks the Ecole Doctorale Astronomie et Astrophysique d’Ile de France for funding their thesis project.
 The authors warmly thank J. Bland-Hawthorn for discussions on surveys and their uncertainties.
\\
This research or product makes use of public auxiliary data provided by ESA/Gaia/DPAC/CU5 and prepared by Carine Babusiaux, for computing the extinction in the Gaia passbands.
\\
This work has made use of data from the European Space Agency (ESA) mission Gaia (\url{https://www.cosmos.esa.int/gaia}), processed by the Gaia Data Processing and Analysis Consortium (DPAC, \url{https://www.cosmos.esa.int/web/gaia/dpac/consortium}). Funding for the DPAC has been provided by national institutions, in particular the institutions participating in the Gaia Multilateral Agreement.
\\
Guoshoujing Telescope (the Large Sky Area Multi-Object Fiber Spectroscopic Telescope LAMOST) is a National Major Scientific Project built by the Chinese Academy of Sciences. Funding for the project has been provided by the National Development and Reform Commission. LAMOST is operated and managed by the National Astronomical Observatories, Chinese Academy of Sciences.
\\
This work made use of the Third Data Release of the GALAH Survey (Buder et al. 2021). The GALAH Survey is based on data acquired through the Australian Astronomical Observatory, under programs: A/2013B/13 (The GALAH pilot survey); A/2014A/25, A/2015A/19, A2017A/18 (The GALAH survey phase 1); A2018A/18 (Open clusters with HERMES); A2019A/1 (Hierarchical star formation in Ori OB1); A2019A/15 (The GALAH survey phase 2); A/2015B/19, A/2016A/22, A/2016B/10, A/2017B/16, A/2018B/15 (The HERMES-TESS program); and A/2015A/3, A/2015B/1, A/2015B/19, A/2016A/22, A/2016B/12, A/2017A/14 (The HERMES K2-follow-up program). We acknowledge the traditional owners of the land on which the AAT stands, the Gamilaraay people, and pay our respects to elders past and present. This paper includes data that has been provided by AAO Data Central (datacentral.org.au).
\\
This work made use of the Fourth Data Release of the GALAH Survey (Buder et al. 2021). The GALAH Survey is based on data acquired through the Australian Astronomical Observatory, under programs: A/2013B/13 (The GALAH pilot survey); A/2014A/25, A/2015A/19, A2017A/18 (The GALAH survey phase 1); A2018A/18 (Open clusters with HERMES); A2019A/1 (Hierarchical star formation in Ori OB1); A2019A/15, A/2020B/23, R/2022B/5, R/2023A/4, R2023B/5 (The GALAH survey phase 2); A/2015B/19, A/2016A/22, A/2016B/10, A/2017B/16, A/2018B/15 (The HERMES-TESS program); A/2015A/3, A/2015B/1, A/2015B/19, A/2016A/22, A/2016B/12, A/2017A/14, A/2020B/14 (The HERMES K2-follow-up program); R/2022B/02 and A/2023A/09 (Combining asteroseismology and spectroscopy in K2); A/2023A/8 (Resolving the chemical fingerprints of Milky Way mergers); and A/2023B/4 (s-process variations in southern globular clusters). We acknowledge the traditional owners of the land on which the AAT stands, the Gamilaraay people, and pay our respects to elders past and present. This paper includes data that has been provided by AAO Data Central (datacentral.org.au).
\\
This work has used data from the APOGEE~DR17 catalogue. Funding for the Sloan Digital Sky Survey IV has been provided by the Alfred P. Sloan Foundation, the U.S. Department of Energy Office of Science, and the Participating Institutions. SDSS acknowledges support and resources from the Center for High-Performance Computing at the University of Utah. The SDSS website is \url{www.sdss4.org}. SDSS is managed by the Astrophysical Research Consortium for the Participating Institutions of the SDSS Collaboration including the Brazilian Participation Group, the Carnegie Institution for Science, Carnegie Mellon University, Center for Astrophysics | Harvard \& Smithsonian (CfA), the Chilean Participation Group, the French Participation Group, Instituto de Astrofísica de Canarias, The Johns Hopkins University, Kavli Institute for the Physics and Mathematics of the Universe (IPMU) / University of Tokyo, the Korean Participation Group, Lawrence Berkeley National Laboratory, Leibniz Institut für Astrophysik Potsdam (AIP), Max-Planck-Institut für Astronomie (MPIA Heidelberg), Max-Planck-Institut für Astrophysik (MPA Garching), Max-Planck-Institut für Extraterrestrische Physik (MPE), National Astronomical Observatories of China, New Mexico State University, New York University, University of Notre Dame, Observatório Nacional / MCTI, The Ohio State University, Pennsylvania State University, Shanghai Astronomical Observatory, United Kingdom Participation Group, Universidad Nacional Autónoma de México, University of Arizona, University of Colorado Boulder, University of Oxford, University of Portsmouth, University of Utah, University of Virginia, University of Washington, University of Wisconsin, Vanderbilt University, and Yale University.
\\
The following python libraries were used for this study: \texttt{Astropy} \citep{astropy_2013, astropy_2018, astropy_2022}, \texttt{scipy} \citep{scipy}, \texttt{numpy} \citep{numpy}, \citep{scikit-learn} and \texttt{matplotlib} \citep{matplotlib}.

\end{acknowledgements}

\bibliographystyle{aa_url}
\bibliography{bibliography.bib}

\begin{appendix}

\section{Neural network architecture}\label{app:architecture}
\begin{figure}[H]
    \centering
    \includegraphics[width=\linewidth]{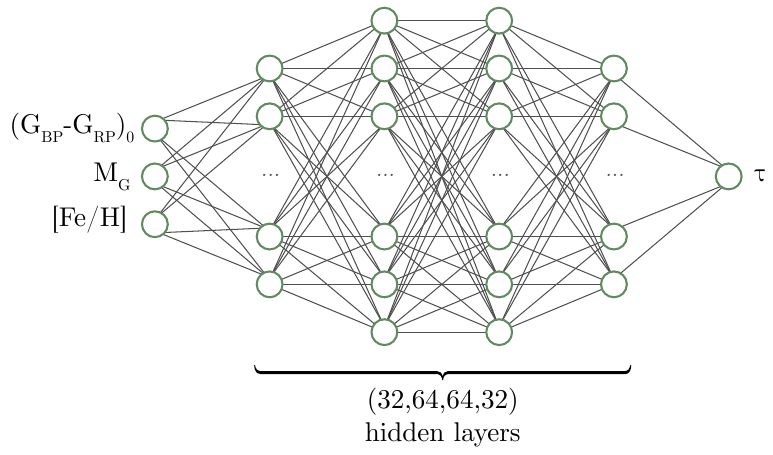}
    \caption{Architecture of the neural networks used in this work.}
    \label{fig:architecture}
\end{figure}
The general architecture of the neural networks used in this work is represented in Fig.~\ref{fig:architecture}.

\section{Effective temperature-colour relation}\label{app:teff_col}
\begin{figure}
    \centering
    \includegraphics[width=\linewidth]{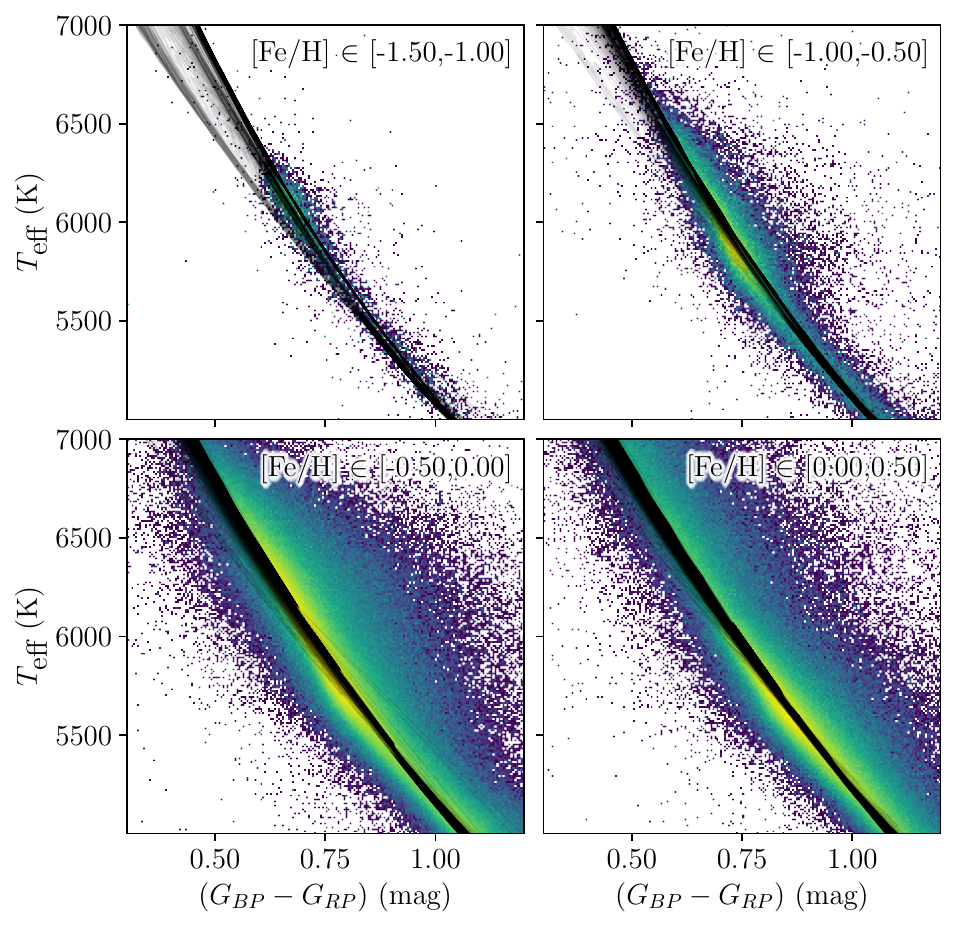}
    \caption{Effective temperature--colour index relation for the BaSTI stellar evolution grid (black curves) and stars from the LAMOST DR10 LRS catalogue (colourmap).}
    \label{fig:teff_col}
\end{figure}

Fig.~\ref{fig:teff_col} shows the effective temperature-colour relation, at different metallicities, for the LAMOST DR10 LRS sample, compared to the theoretical relation given by the BaSTI grid. The colour index is given by Gaia photometry $(G_{BP}-G_{RP})$ and the effective temperature comes from the LAMOST catalogue. While the observational relation matches the theoretical one without any offset, in all metallicity bins, the spread is significant, meaning a non-negligible part of our sample has an estimated effective temperature that differs from the one our theoretical grid expects given the same colour index. At lower metallicities, some of the other catalogues can also display offsets from the theoretical relations (e.g. the GALAH DR4 catalogue at metallicities lower than 0.3).

{Dependence of age estimations on Monte Carlo samples}\label{app:MC_samples}

\begin{figure}
    \centering
    \includegraphics[width=\linewidth]{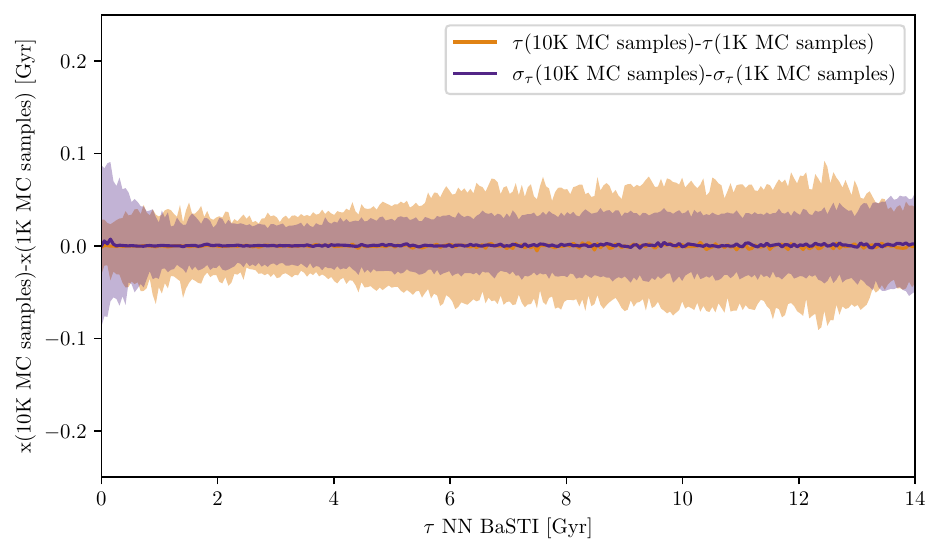}
    \caption{Age versus difference in median age (orange) and standard deviation (purple) estimates, using 1000 and 10000 Monte Carlo samples. Shaded areas show the 16th and 84th percentiles.}
    \label{fig:MC_samples}
\end{figure}

The choice of number of Monte Carlo samples to estimate the age PDF of a star is arbitrary, with a higher number resulting in a better convergence towards the true age distribution, as well as a higher computational cost. In Fig.~\ref{fig:MC_samples}, we compare the age medians and associated standard deviations of GALAH DR4 stars, when using 10 000 and 1000 Monte Carlo samples. We find that the median difference of median ages, marginalized over all ages, is of $-3.3*10^{-5}\pm 0.06$ Gyr, while the median difference of standard deviations, marginalized over all ages, is of $4.8*10^{-4}\pm 0.03$ Gyr. Thus, for the vast majority of cases, 1000 Monte Carlo samples is enough to faithfully represent the underlying true age PDF, whereas some extreme cases necessitate more samples.

\section{Evolutionary model parameters}\label{app:models}
The different parameter spaces of each model used in this work are given in Table~\ref{tab:models}. We note that the number of samples taken for metallicities, ages and masses are not the same in the different models, leading to a different number of points in each grid. The provided ranges of parameters help compare the domains of applicability of each model.

\begin{table}[H]
    \caption{Parameter spaces for each stellar evolution model.}
    \label{tab:models}      
    \centering          
    \begin{tabular}{c | c c c }
        Model name & [M/H] & Age (Gyr) & M ($M_\odot$) \\
        \hline
        BaSTI & [-3.2,0.3] & [$5\times10^{-6}$,14] & [0.1,10] \\
        MIST & [-5.0,0.5] & [0.1,12.6] & [0.1,5.2] \\
        PARSEC & [-0.9,0.3] & [0.1,13.9] & [0.1,5] \\
        Dartmouth & [-2.6,0.7] & [0.3,13.5] & [0.1,3.8] \\
        Geneva & [-1.5,0.2] & [$1\times10^{-2}$,12.6] & [0.8,500] \\
    \end{tabular}
\end{table}

\section{Comparison of different grid age estimates using GALAH DR4 stars}\label{app:model_comp}

\begin{figure*}
    \centering
    \includegraphics[width=\linewidth]{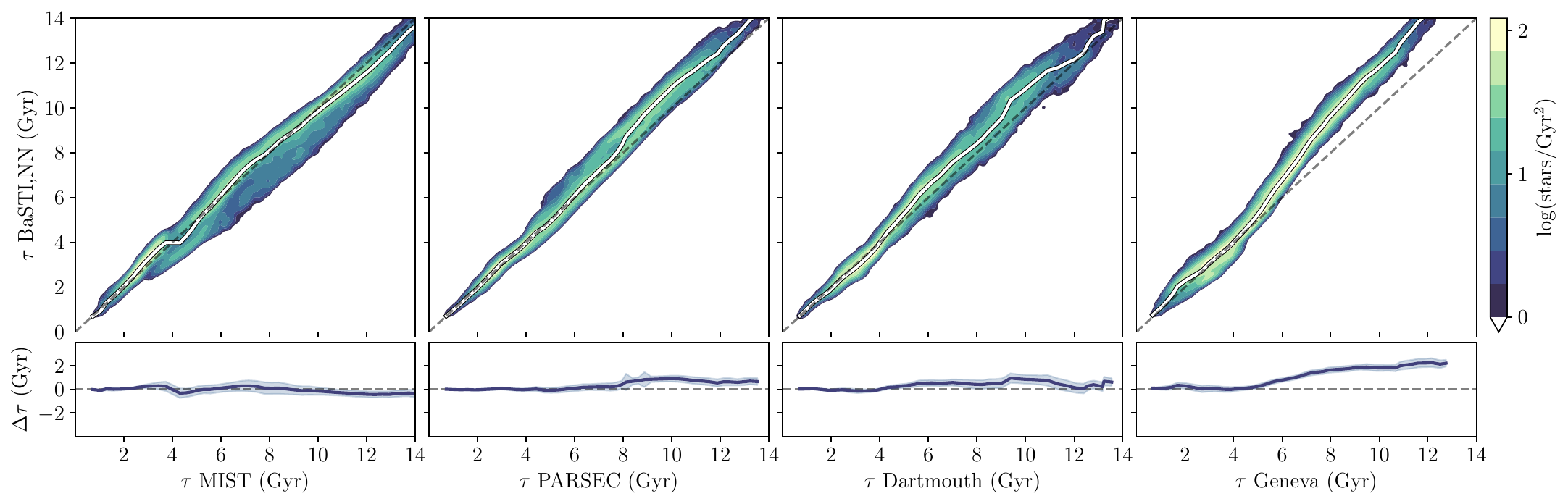}
    \caption{\textit{Top panels:} Age estimates of GALAH DR4 stars from the NN trained on BaSTI compared to those from NNs trained on (from left to right) MIST, PARSEC, Dartmouth and SYCLIST evolutionary grids. Running medians are shown in white lines, as well as a one-to-one black dashed line to guide the eye. \textit{Bottom panels:} Running median of the difference $\Delta\tau$ between the age estimates of our BaSTI-trained NN and other NNs in solid lines, and associated standard deviations in shades areas.}
    \label{fig:model_comp_galah}
    \vspace{0.0cm}
\end{figure*}

In this section, we compare age estimates made using neural networks trained on different stellar evolutionary models, as was done in Fig.~\ref{fig:model_comp}, but this time using GALAH DR4 stars instead of C24's dataset. The resulting distributions are plotted in Fig.~\ref{fig:model_comp_galah}, where age estimates made with a BaSTI-trained neural network are compared to age estimates made with neural networks trained on the MIST, PARSEC, Dartmouth and Geneva grids. We find that BaSTI ages still agree with MIST ages over the whole age range. For Dartmouth ages, the standard deviation is greatly reduced from a median value over the whole age range of 0.40 Gyr for C24's dataset to 0.27 Gyr for GALAH DR4. The highest deviation is seen at ages $\sim$ 10 Gyr, that was not seen on C24's dataset, although still compatible with it within 1 $\sigma$. For Geneva ages, the trends are similar with GALAH DR4 and C24 datasets, although the former displays lower standard deviations, from a median value of 0.31 Gyr to 0.22 Gyr. The PARSEC ages show the biggest difference with C24's dataset. With GALAH DR4 stars, the BaSTI and PARSEC ages show a better agreement, with very little offset for ages < 8 Gyr. For older ages, the difference is of about 1 Gyr, and 0.5 Gyr for ages > 12 Gyr. The standard deviation is also lower, with a median value of 0.26 compared to 0.52 Gyr using C24's dataset. The difference in grid comparisons can mainly be explained by the CMD coverage of the sample used. In the case of C24's dataset, used in Fig.~\ref{fig:model_comp}, older stars are distributed closer to the Zero Age Main Sequence, where isochrone degeneracy is high, while in the case of GALAH DR4 stars, used here, older stars tend to populate further regions along those isochrones, where degeneracy is lower.

\section{Spectroscopic surveys used in this work}\label{app:surveys}

\begin{figure*}
    \centering
    \includegraphics[width=\linewidth]{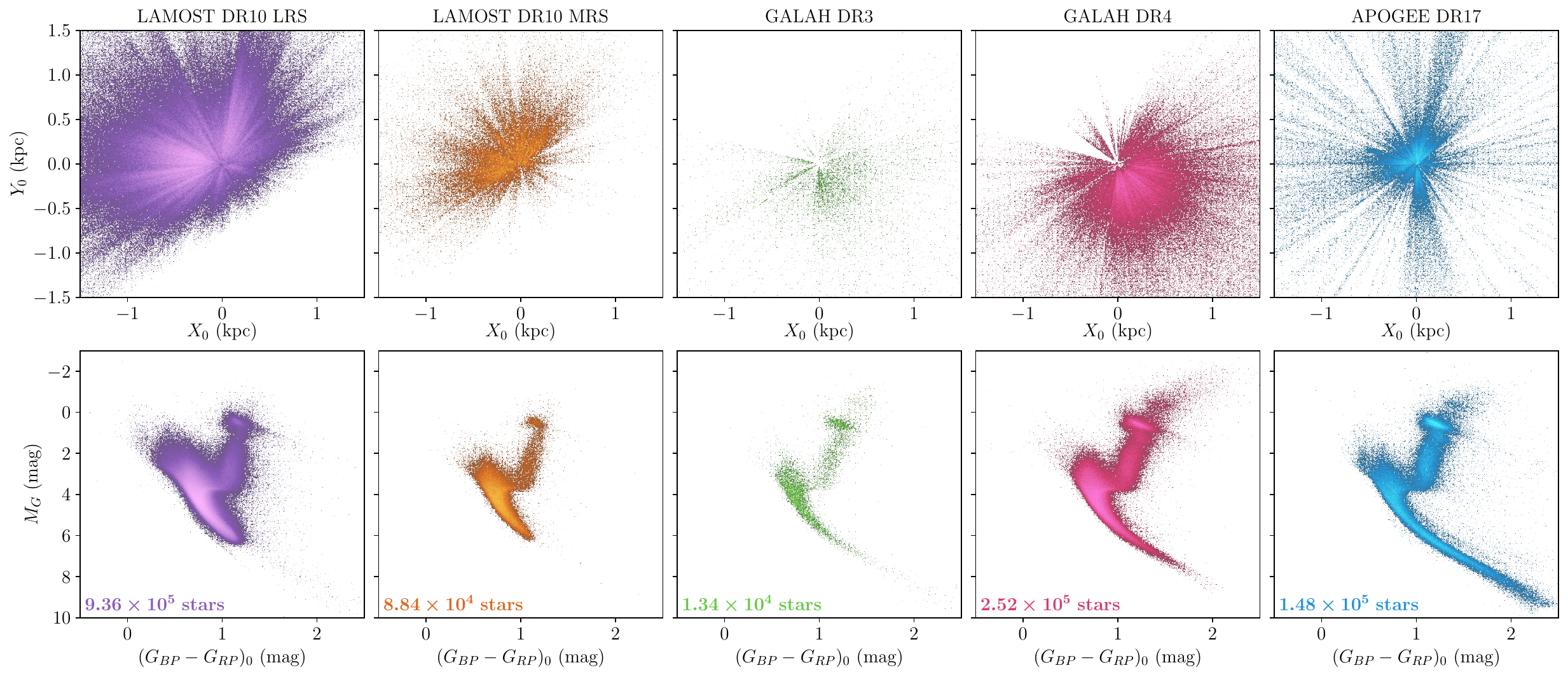}
    \caption{Overview of the surveys used in this work, coloured by density. From left to right: LAMOST DR10 \citep{Lamost_cui,Lamost_zhao}, LRS (purple) and MRS (orange), GALAH DR3 \citep{GALAH_dr3_De_silva,Galah_DR3} (green), GALAH DR4 \citep{GALAH_dr3_De_silva,Galah_DR4} (pink) and APOGEE DR17 \citep{APOGEE} (blue). \textit{Top panel:} Face-on ($X_0$,$Y_0$) galactic maps, centred on the Sun. \textit{Bottom panel:} Hertzsprung-Russell Diagrams, coloured by density. The number of stars in each sample is annotated in the bottom left panel corner. Lighter colours correspond to higher density regions.}
    \label{fig:hr_xy}
    \vspace{0.0cm}
\end{figure*}

The samples defined in this work based on several spectroscopic surveys are displayed in a ($X_0$,$Y_0$) Galactic face-on view and as CMDs in Fig.~\ref{fig:hr_xy}. Each sample covers a wide range of the MS, MSTO, Red Giant Branch (RGB) and SGB phases, with varying degrees of spread. Our LAMOST DR10 LRS sample offers the most stars, but it also has the most spread out CMD (as a consequence of the high number of stars covering more regions, but also because of the bigger spatial coverage, reaching distances of higher extinctions, where extinction maps are less precise), while the GALAH DR4 and APOGEE DR17 samples offer more thin sequences. Together, the samples cover a wide and diverse region of the solar neighbourhood. In all samples, the most densely populated regions of the CMD are the lower-magnitude end of the MS, and the MSTO, along with a notable over-density in the red clump.

\section{Influence of $A_V$ on age estimations}\label{app:av_comp}

\begin{figure}
    \centering
    \includegraphics[width=.99\linewidth]{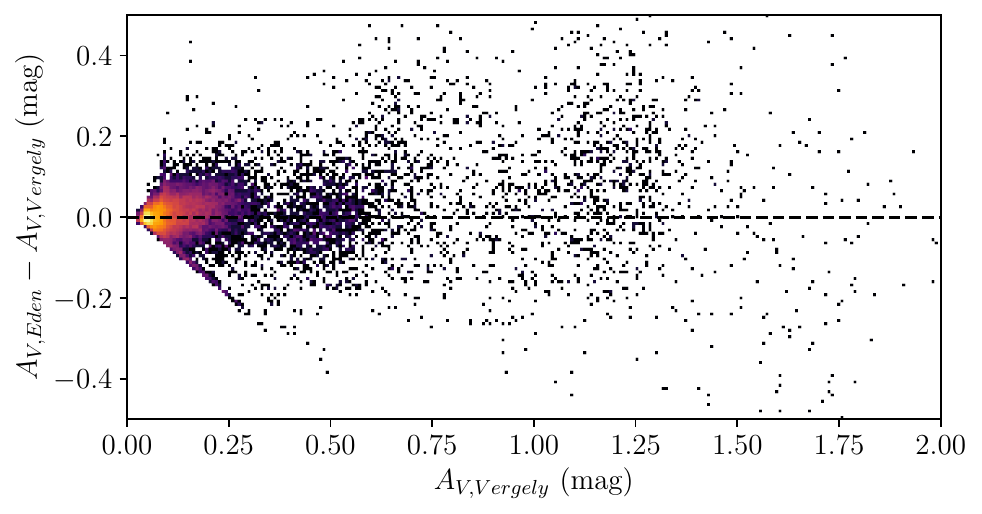}
    \includegraphics[width=.99\linewidth]{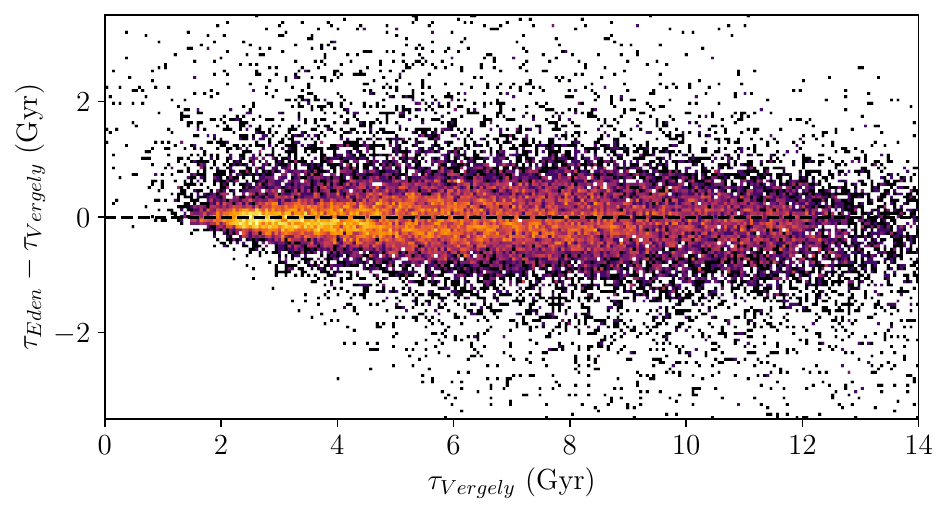}
    \caption{Comparison of $A_V$ (top panel) values from \cite{vergely} and \cite{edenhofer}. From these values, different extinction corrections are applied, which give different age estimations (bottom panel).}
    \label{fig:Av_comp}
\end{figure}

Stellar extinction effects offset the observed position of a star in the CMD compared to its true position. Thus, age estimates based on this position have to rely on precise $A_V$ values. To demonstrate this effect, we plot in Fig.~\ref{fig:Av_comp} age determinations based on the same observed properties, but corrected for extinction using two different extinction maps (one from \cite{vergely} and one from \cite{edenhofer}), for stars of C24's sample.

The top panel of Fig.~\ref{fig:Av_comp} shows the difference in $A_V$ values computed for individual stars by integrating the extinction maps along the line of sight from the star to the sun, when the extinction maps used are different. This, in turn, will affect the age estimation, which is shown in the bottom panel of Fig.~\ref{fig:Av_comp}. In some cases, the effect of a different value of $A_V$ can lead to several Gyr of difference, highlighting the need for precise extinction maps for stellar age estimation. We note that no error on this extinction value is provided by both maps used here.

\section{Age-abundance planes for APOGEE DR17 and different evolutionary models}\label{app:age_abund}

\begin{figure*}
    \centering
    \includegraphics[width=\linewidth]{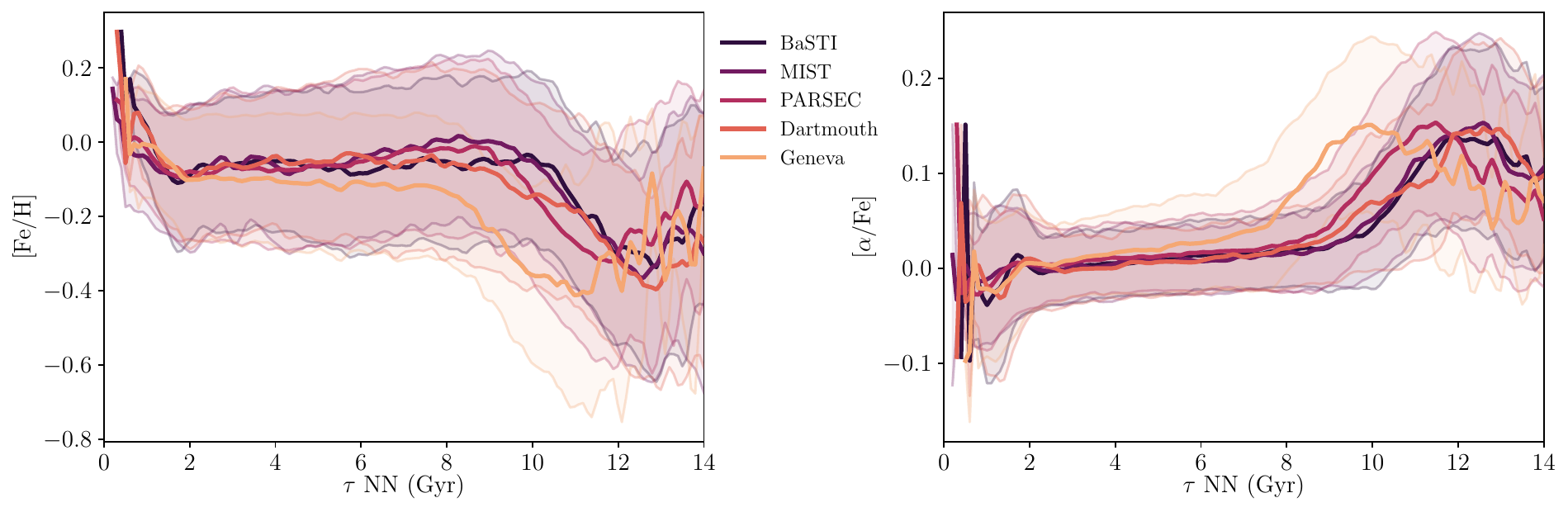}
    \caption{Abundance versus age trends for our APOGEE DR17 sample, with age estimated using NN trained on the BaSTI, MIST, PARSEC, Dartmouth, and Geneva grids. Coloured lines and shaded areas are the means and $\pm1\sigma$ of the distributions.}
    \label{fig:age_abundance_models}
\end{figure*}

In this section, we compare the AMRs and [$\alpha$/Fe]-age relations given for our APOGEE DR17 sample, and with ages estimated using our NNs trained on different evolutionary models. As was discussed in Section~\ref{sec:models}, the BaSTI, MIST, PARSEC and Dartmouth trained NNs produce similar age estimates, and thus similar age-abundance relations. The two sequences of flat/shallow slopes at ages < 10 Gyr and steeper slopes at older ages, both seen in the AMR (left subplot of Fig.~\ref{fig:age_abundance_models}) and the [$\alpha$/Fe]-age relation (right subplot), are similar. We observe a slight deviation at older ages in the [$\alpha$/Fe]-age relation for the PARSEC-trained NNs, with a peak at ages $\sim$ 11 Gyr, while the peak is at ages $\sim$ 12 Gyr for the BaSTI and Dartmouth-trained NNs. The Geneva-trained NN gives different age estimates, and the corresponding relations diverge from the others, especially at older ages, with a drop/peak positioned at younger ages $\sim$ 9-10 Gyr.

\section{Influence of $\alpha$ elements on age estimations}\label{app:alpha_comp}

\begin{figure*}
    \centering
    \includegraphics[width=\linewidth]{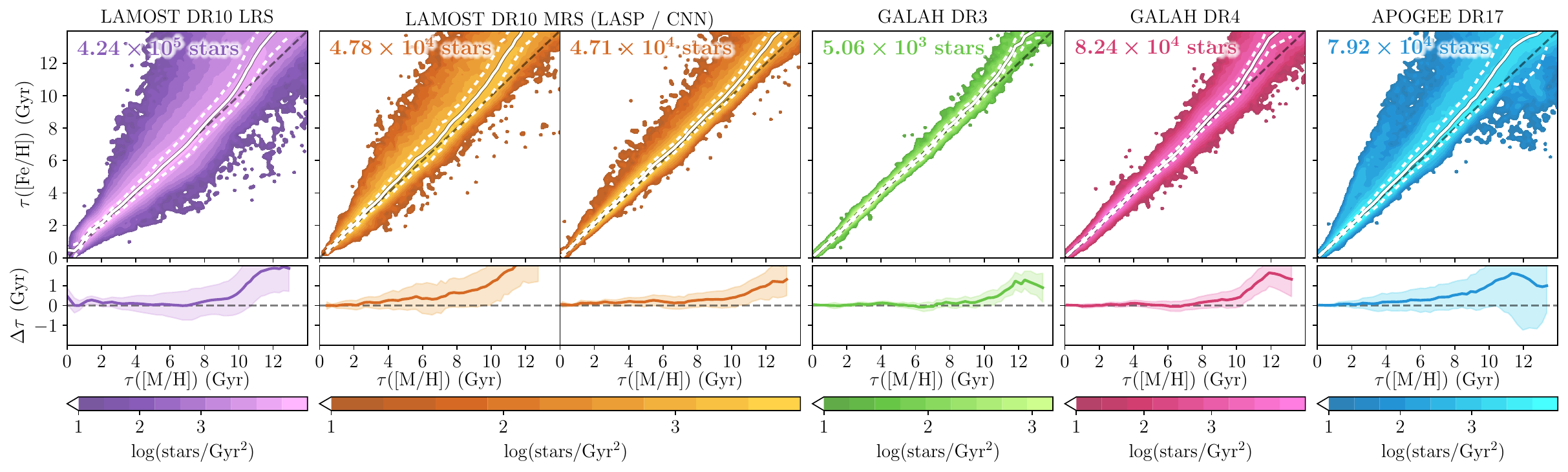}
    \caption{Age comparison with scaled ([M/H]) and unscaled ([Fe/H]) metallicities, from anNN trained on the BaSTI grid. \textit{Top panel:} Age from scaled metallicity $\tau$([M/H]) versus age from unscaled metallicity $\tau$([Fe/H]). The number of stars in each sample is annotated in the upper panel corners, and running medians and $\pm\sigma$ lines are shown in solid and dotted white lines, as well as a one-to-one black dashed line to guide the eye. \textit{Bottom panel:} Running median of the difference $\Delta\tau$ between the two age estimates in solid lines, and associated standard deviation in shaded areas.}
    \label{fig:app_age_comp_alpha}
\end{figure*}

In this section, we compare age estimates made with a metallicities scaled using \cite{Salaris} relation (see Section~\ref{sec:data}) and with those that do not use such scaling relation. We estimate the ages for the LAMOST, GALAH and APOGEE catalogues, and compare them in Fig~\ref{fig:app_age_comp_alpha}. The GALAH catalogues show a very tight relation with low dispersion, and offset up to ages $\sim$ 10 Gyr, whereas LAMOST DR10 and APOGEE DR17 show a similar agreement for this age range, but with higher dispersion. The APOGEE DR17 panel seems to show a secondary sequence seen between 0 and 6 Gyr, for which unscaled metallicity ages reach ages higher than 10 Gyr. For ages older than 10 Gyr, all catalogues show a bump of greater ages for the unscaled-metallicity ages. This behaviour is expected as, for ages < 10 Gyr, the $\alpha$ element abundance distribution is centred on 0.0, while for older ages, the thick-disk [$\alpha$/Fe] increases with age (see Fig.~\ref{fig:age_abundance} and Section~\ref{sec:abund}). Hence, for young stars, on average, we have:
$$
\text{[M/H]} \approx \text{[Fe/H]}
$$
On the other hand, for an average old star having [$\alpha$/Fe] $\approx$ 0.2, we have:
$$
\text{[M/H]} \approx \text{[Fe/H]} + 0.76*0.2 = \text{[Fe/H] + 0.152}
$$
This increase in the scaled metallicity will generally lead to a decrease in the age estimate, explaining the observed trends in Fig.~\ref{fig:app_age_comp_alpha}. The age at which $\Delta\tau$ begins to differ is thus a proxy for the start of the thick disk sequence.

\end{appendix}
\label{LastPage}
\end{document}